\documentclass[twocolumn,tighten]{aastex701}
\usepackage[utf8]{inputenc}
\usepackage[T1]{fontenc}
\usepackage{CJKutf8}
\usepackage{float}
\usepackage{amsmath}
\usepackage{enumitem}
\usepackage{booktabs,pbox}
\usepackage{bm}
\usepackage{colortbl,multirow}

\graphicspath{{./figures/}}

\newcommand{\sbullet}{\mathbin{\vcenter{\hbox{\scalebox{0.7}{$\bullet$}}}}}
\newcommand{\sbsc}[1]{_\mathrm{#1}}
\newcommand{\spsc}[1]{^\mathrm{#1}}

\newcommand{\HI}{\ion{H}{1}}

\newcommand{\Htwo}{\mbox{$\mathrm{H}_2$}}
\newcommand{\CI}{\ion{C}{1}}
\newcommand{\CII}{\ion{C}{2}}
\newcommand{\CO}[2]{\mbox{$\mathrm{CO}\,(#1\text{--}#2)$}}

\newcommand{\SigSFR}{\Sigma\sbsc{SFR}}
\newcommand{\Sigstar}{\Sigma_\star}

\newcommand{\radius}[1]{r\sbsc{#1}}
\newcommand{\SFR}{\mathrm{SFR}}
\newcommand{\Mstar}{M_\star}
\newcommand{\DeltaMS}{\Delta\mathrm{MS}}
\newcommand{\Mmol}{M\sbsc{mol}}
\newcommand{\LCO}{L^\prime\sbsc{CO}}
\newcommand{\LCOxy}[2]{L^\prime\sbsc{CO(#1{-}#2)}}
\newcommand{\tdep}{t\sbsc{dep}}

\newcommand{\ICOxy}[2]{I\sbsc{CO(#1{-}#2)}}
\newcommand{\alphaCO}{\alpha\sbsc{CO}}
\newcommand{\alphaCOxy}[2]{\alpha\sbsc{CO(#1{-}#2)}}

\newcommand{\uI}{\mathrm{MJy\;sr^{-1}}}

\newcommand{\uIco}{\mathrm{K\;km\;s^{-1}}}
\newcommand{\uLco}{\mathrm{K\;km\;s^{-1}\;pc^2}}
\newcommand{\ualphaCO}{\mathrm{M_\odot\;pc^{-2}\;(K\;km\;s^{-1})^{-1}}}

\newcommand{\uM}{\mathrm{M_\odot}}
\newcommand{\uSig}{\mathrm{M_\odot\;pc^{-2}}}

\newcommand{\uSFR}{\mathrm{M_\odot\;yr^{-1}}}
\newcommand{\uSigSFR}{\mathrm{M_\odot\;yr^{-1}\;kpc^{-2}}}

\newcommand{\ngalall}{10,657}

\newcommand{\ngalstarsfr}{5,300}
\newcommand{\ngalstarsfrrp}{3,956}
\newcommand{\ngalxco}{5,244}

\defcitealias{Schinnerer_Leroy_2024}{SL24}
\defcitealias{Leroy_etal_2019}{L19}

\shorttitle{CO-to-H$_2$ Conversion Factor across Thousands of Galaxies}
\shortauthors{Sun et al.}

\begin{document}

\title{Resolved Profiles of Stellar Mass, Star Formation Rate, and Predicted CO-to-H$_2$ Conversion Factor Across Thousands of Local Galaxies}


\newcommand{\Kentucky}{Department of Physics and Astronomy, University of Kentucky, 506 Library Drive, Lexington, KY 40506, USA}

\newcommand{\Princeton}{Department of Astrophysical Sciences, Princeton University, 4 Ivy Lane, Princeton, NJ 08544, USA}

\newcommand{\McMaster}{Department of Physics and Astronomy, McMaster University, 1280 Main Street West, Hamilton, ON L8S 4M1, Canada}

\newcommand{\CITA}{Canadian Institute for Theoretical Astrophysics (CITA), University of Toronto, 60 St George Street, Toronto, ON M5S 3H8, Canada}

\newcommand{\OSU}{Department of Astronomy, The Ohio State University, 140 West 18th Avenue, Columbus, OH 43210, USA}

\newcommand{\CCAPP}{Center for Cosmology and Astroparticle Physics (CCAPP), 191 West Woodruff Avenue, Columbus, OH 43210, USA}

\newcommand{\Alberta}{Department of Physics, University of Alberta, Edmonton, AB T6G 2E1, Canada}

\newcommand{\ANU}{Research School of Astronomy and Astrophysics, Australian National University, Canberra, ACT 2611, Australia}

\newcommand{\Arizona}{Steward Observatory, University of Arizona, 933 North Cherry Avenue, Tucson, AZ 85721, USA}

\newcommand{\ASIAA}{Institute of Astronomy and Astrophysics, Academia Sinica, No. 1, Sec. 4, Roosevelt Road, Taipei 106216, Taiwan}

\newcommand{\ASTROThreeD}{ARC Centre of Excellence for All Sky Astrophysics in 3 Dimensions (ASTRO 3D), Australia}

\newcommand{\Bonn}{Argelander-Institut f\"ur Astronomie, Universit\"at Bonn, Auf dem H\"ugel 71, 53121 Bonn, Germany}

\newcommand{\Carnegie}{Observatories of the Carnegie Institution for Science, 813 Santa Barbara Street, Pasadena, CA 91101, USA}

\newcommand{\CCA}{Center for Computational Astrophysics, Flatiron Institute, 162 Fifth Avenue, New York, NY 10010, USA}

\newcommand{\CfA}{Center for Astrophysics $\mid$ Harvard \& Smithsonian, 60 Garden Street, Cambridge, MA 02138, USA}

\newcommand{\CITEVA}{Centro de Astronomía (CITEVA), Universidad de Antofagasta, Avenida Angamos 601, Antofagasta, Chile}

\newcommand{\CNRS}{CNRS, IRAP, 9 Av. du Colonel Roche, BP 44346, F-31028 Toulouse cedex 4, France}

\newcommand{\COOL}{Cosmic Origins Of Life (COOL) Research DAO, coolresearch.io}

\newcommand{\EPFL}{Institute of Physics, Laboratory for galaxy evolution and spectral modelling, EPFL, Observatoire de Sauverny, Chemin Pegais 51, 1290 Versoix, Switzerland}

\newcommand{\ESO}{European Southern Observatory, Karl-Schwarzschild Stra{\ss}e 2, D-85748 Garching bei M\"{u}nchen, Germany}

\newcommand{\Gent}{Sterrenkundig Observatorium, Universiteit Gent, Krijgslaan 281 S9, B-9000 Gent, Belgium}

\newcommand{\Hawaii}{Institute for Astronomy, University of Hawaii, 2680 Woodlawn Drive, Honolulu, HI 96822, USA}

\newcommand{\Heidelberg}{Astronomisches Rechen-Institut, Zentrum f\"{u}r Astronomie der Universit\"{a}t Heidelberg, M\"{o}nchhofstra\ss e 12-14, D-69120 Heidelberg, Germany}

\newcommand{\IAC}{Instituto de Astrof\'isica de Canarias, C/ V\'ia L\'actea s/n, E-38205, La Laguna, Spain}

\newcommand{\IALP}{Instituto de Astrofísica de La Plata, CONICET--UNLP, Paseo del Bosque S/N, B1900FWA La Plata, Argentina}

\newcommand{\ICRAR}{International Centre for Radio Astronomy Research, University of Western Australia, 35 Stirling Highway, Crawley, WA 6009, Australia}

\newcommand{\INAF}{INAF -- Osservatorio Astrofisico di Arcetri, Largo E. Fermi 5, I-50157, Firenze, Italy}

\newcommand{\IPAC}{Caltech-IPAC, 1200 E. California Blvd. Pasadena, CA 91125, USA}

\newcommand{\IPARC}{Instituto de F\'{\i}sica de Part\'{\i}culas y del Cosmos IPARCOS, Facultad de Ciencias F\'{\i}sicas, Universidad Complutense de Madrid, E-28040, Spain}

\newcommand{\IRAM}{Institut de Radioastronomie Millim\'etrique (IRAM), 300 Rue de la Piscine, F-38406 Saint Martin d'H\`eres, France}

\newcommand{\ITA}{Universit\"{a}t Heidelberg, Zentrum f\"{u}r Astronomie, Institut f\"{u}r Theoretische Astrophysik, Albert-Ueberle-Str 2, D-69120 Heidelberg, Germany}

\newcommand{\IWR}{Universit\"{a}t Heidelberg, Interdisziplin\"{a}res Zentrum f\"{u}r Wissenschaftliches Rechnen, Im Neuenheimer Feld 205, D-69120 Heidelberg, Germany}

\newcommand{\JHU}{Department of Physics and Astronomy, The Johns Hopkins University, Baltimore, MD 21218, USA}

\newcommand{\Kansas}{Department of Physics and Astronomy, University of Kansas, 1251 Wescoe Hall Drive, Lawrence, KS 66045, USA}

\newcommand{\LAM}{Aix Marseille Univ, CNRS, CNES, LAM (Laboratoire d’Astrophysique de Marseille), Marseille, France}

\newcommand{\Leiden}{Leiden Observatory, Leiden University, P.O. Box 9513, 2300 RA Leiden, The Netherlands}

\newcommand{\Liverpool}{Astrophysics Research Institute, Liverpool John Moores University, IC2, Liverpool Science Park, 146 Brownlow Hill, Liverpool L3 5RF, UK}

\newcommand{\Lyon}{Univ Lyon, Univ Lyon 1, ENS de Lyon, CNRS, Centre de Recherche Astrophysique de Lyon UMR5574, F-69230 Saint-Genis-Laval, France}

\newcommand{\UMD}{Department of Astronomy, University of Maryland, 4296 Stadium Drive, College Park, MD 20742, USA}

\newcommand{\MPE}{Max-Planck-Institut f\"{u}r extraterrestrische Physik, Giessenbachstra{\ss}e 1, D-85748 Garching, Germany}

\newcommand{\MPIA}{Max-Planck-Institut f\"{u}r Astronomie, K\"{o}nigstuhl 17, D-69117, Heidelberg, Germany}

\newcommand{\Nagoya}{Department of Physics, Nagoya University, Furo-cho, Chikusa-ku, Nagoya, Aichi 464-8602, Japan}

\newcommand{\NAOJ}{National Astronomical Observatory of Japan, 2-21-1 Osawa, Mitaka, Tokyo, 181-8588, Japan}

\newcommand{\Nichidai}{Department of Physics, General Studies, College of Engineering, Nihon University, 1 Nakagawara, Tokusada, Tamuramachi, Koriyama, Fukushima, 963-8642, Japan}

\newcommand{\NRAO}{National Radio Astronomy Observatory, 520 Edgemont Road, Charlottesville, VA 22903, USA}

\newcommand{\OAN}{Observatorio Astron\'{o}mico Nacional (IGN), C/Alfonso XII, 3, E-28014 Madrid, Spain}

\newcommand{\ObsParis}{Sorbonne Universit\'{e}, Observatoire de Paris, Universit\'{e} PSL, CNRS, LERMA, F-75014, Paris, France}

\newcommand{\Oxford}{Sub-department of Astrophysics, Department of Physics, University of Oxford, Keble Road, Oxford OX1 3RH, UK}

\newcommand{\Rutgers}{Department of Physics and Astronomy, Rutgers, the State University of New Jersey, 136 Frelinghuysen Road, Piscataway, NJ 08854, USA}

\newcommand{\STScI}{Space Telescope Science Institute, 3700 San Martin Drive, Baltimore, MD 21218, USA}

\newcommand{\STScIESA}{AURA for the European Space Agency (ESA), Space Telescope Science Institute, 3700 San Martin Drive, Baltimore, MD 21218, USA}

\newcommand{\Surrey}{Department of Physics, University of Surrey, Guildford GU2 7XH, UK}

\newcommand{\Sydney}{Sydney Institute for Astronomy, School of Physics A28, The University of Sydney, NSW 2006, Australia}

\newcommand{\TAPIR}{TAPIR, California Institute of Technology, Pasadena, CA 91125, USA}

\newcommand{\Tamkang}{Department of Physics, Tamkang University, No.151, Yingzhuan Rd., Tamsui Dist., New Taipei City 251301, Taiwan}

\newcommand{\Toulouse}{Universit\'{e} de Toulouse, UPS-OMP, IRAP, F-31028 Toulouse cedex 4, France}

\newcommand{\Toledo}{University of Toledo, 2801 W. Bancroft St., Mail Stop 111, Toledo, OH 43606, USA}

\newcommand{\UChile}{Departamento de Astronom\'{i}a, Universidad de Chile, Camino del Observatorio 1515, Las Condes, Santiago, Chile}

\newcommand{\UCM}{Departamento de F\'{\i}sica de la Tierra y Astrof\'{\i}sica, Universidad Complutense de Madrid, E-28040, Spain}

\newcommand{\UCSD}{Department of Astronomy \& Astrophysics,  University of California, San Diego, 9500 Gilman Drive, La Jolla, CA 92093, USA}

\newcommand{\ULL}{Departamento de Astrof\'isica, Universidad de La Laguna, Av. del Astrof\'isico Francisco S\'anchez s/n, E-38206, La Laguna, Spain}

\newcommand{\UMass}{University of Massachusetts—Amherst, 710 North Pleasant Street, Amherst, MA 01003, USA}

\newcommand{\UVa}{University of Virginia, 530 McCormick Road, Charlottesville, VA 22904, USA}

\newcommand{\Wyoming}{Department of Physics and Astronomy, University of Wyoming, Laramie, WY 82071, USA}

\newcommand{\Zurich}{Institute for Computational Science, University of Z\"urich, Winterthurerstrasse 190, 8057 Z\"urich, Switzerland}


\author[0000-0003-0378-4667]{Jiayi~Sun \textnormal{\begin{CJK*}{UTF8}{gbsn}(孙嘉懿)\end{CJK*}}}
\altaffiliation{NASA Hubble Fellow}
\affiliation{\Kentucky}
\affiliation{\Princeton}
\email{jysun.princeton@gmail.com}

\author[0000-0003-4209-1599]{Yu-Hsuan~Teng}
\affiliation{\UMD}
\email{yhteng@umd.edu}

\author[0000-0003-2551-7148]{I-Da~Chiang \textnormal{\begin{CJK*}{UTF8}{bkai}(江宜達)\end{CJK*}}}
\affiliation{\ASIAA}
\email{idchiang@asiaa.sinica.edu.tw}

\author[0000-0002-2545-1700]{Adam~K.~Leroy}
\affiliation{\OSU}
\affiliation{\CCAPP}
\email{leroy.42@osu.edu}

\author[0000-0002-4378-8534]{Karin~Sandstrom}
\affiliation{\UCSD}
\email{karin.sandstrom@gmail.com}

\author[0000-0002-8760-6157]{Jakob~den~Brok}
\affiliation{\CfA}
\email{jakob.den_brok@cfa.harvard.edu}

\author[0000-0002-5480-5686]{Alberto~D.~Bolatto}
\affiliation{\UMD}
\email{bolatto@umd.edu}

\author[0000-0002-5235-5589]{J\'er\'emy~Chastenet}
\affiliation{\Gent}
\email{jeremy.chastenet@ugent.be}

\author[0000-0001-8241-7704]{Ryan~Chown}
\affiliation{\OSU}
\email{chown.5@osu.edu}

\author[0000-0002-9181-1161]{Annie~Hughes}
\affiliation{\CNRS}
\email{annie.hughes@irap.omp.eu}

\author[0000-0001-9605-780X]{Eric~W.~Koch}
\affiliation{\CfA}
\email{koch.eric.w@gmail.com}

\author[0000-0002-0012-2142]{Thomas~G.~Williams}
\affiliation{\Oxford}
\email{thomas.williams@physics.ox.ac.uk}



\begin{abstract}
We present radial profiles of surface brightness in UV and IR bands, estimate stellar mass surface density ($\Sigma_\star$) and star formation rate surface density ($\Sigma_\mathrm{SFR}$), and predict the CO-to-H$_2$ conversion factor ($\alpha_\mathrm{CO}$) for over 5,000 local galaxies with stellar mass $M_\star\,{\geq}\,10^{9.3}\rm\,M_\odot$.
We build these profiles and measure galaxy half-light radii using GALEX and WISE images from the $z$0MGS program, with special care given to highly inclined galaxies.
From the UV and IR surface brightness profiles, we estimate $\Sigma_\star$ and $\Sigma_\mathrm{SFR}$ and use them to predict $\alpha_\mathrm{CO}$ with state-of-the-art empirical prescriptions.
We validate our (kpc-scale) $\alpha_\mathrm{CO}$ predictions against observational estimates, finding the best agreement when accounting for CO-dark gas as well as CO emissivity and excitation effects.
The CO-dark correction plays a primary role in lower-mass galaxies, whereas CO emissivity and excitation effects become more important in higher-mass and more actively star-forming galaxies, respectively.
We compare our estimated $\alpha_\mathrm{CO}$ to observed galaxy-integrated SFR to CO luminosity ratio as a function of $M_\star$.
A large compilation of literature data suggests that star-forming galaxies with $M_\star = 10^{9.5{-}11}\,M_\odot$ show strong anti-correlations of SFR/$L^\prime_\mathrm{CO(1{-}0)} \propto M_\star^{-0.29}$ and SFR/$L^\prime_\mathrm{CO(2{-}1)} \propto M_\star^{-0.40}$.
The estimated $\alpha_\mathrm{CO}$ trends, when combined with a constant molecular gas depletion time $t_\mathrm{dep}$, can only explain ${\approx}1/3$ of these SFR/$L^\prime_\mathrm{CO}$ trends.
This suggests that $t_\mathrm{dep}$ being systematically shorter in lower-mass star-forming galaxies is the main cause of the observed SFR/$L^\prime_\mathrm{CO}$ variations.
We publish all data products from this work, including galaxy sizes, UV and IR surface brightness profiles, $\Sigma_\star$, $\Sigma_\mathrm{SFR}$, and $\alpha_\mathrm{CO}$ estimates.
\end{abstract}


\section{Introduction}
\label{sec:intro}

Low-$J$ CO rotational transition lines are among the most widely used tracers of molecular gas in galaxies \citep[see reviews by][]{Young_Scoville_1991,Fukui_Kawamura_2010,Heyer_Dame_2015,Saintonge_Catinella_2022}.
In the local universe, integrated CO measurements exist for thousands of galaxies \citep[e.g.,][]{Young_etal_1995,Lisenfeld_etal_2011,Cicone_etal_2017,Saintonge_etal_2017,Colombo_etal_2020,Wylezalek_etal_2022}, 
resolved CO observations cover hundreds of galaxies \citep[e.g.,][]{Helfer_etal_2003,Kuno_etal_2007,Leroy_etal_2009,Bolatto_etal_2017,Sorai_etal_2019,LinLH_etal_2020,Brown_etal_2021}, 
and there are now over 100 high-resolution CO maps that isolate or even resolve individual molecular clouds across galaxies \citep[e.g.,][J.~Sun et al., in preparation]{DonovanMeyer_etal_2013,Leroy_etal_2021a,Williams_etal_2023}.
These observations have shaped our understanding of the overall abundance, large-scale distribution, and small-scale organization of  cold molecular gas, which is the direct fuel for star formation and a key driver of galaxy evolution.

As the collective footprint of CO observations has grown, our understanding of how to infer molecular gas mass from CO emission has also improved \citep[see][]{Bolatto_etal_2013,Schinnerer_Leroy_2024}.
Many empirical studies made key advances by comparing CO emission to independent gas mass tracers, including far-IR dust emission \citep[e.g.,][]{Leroy_etal_2011,Sandstrom_etal_2013,Yasuda_etal_2023,Chiang_etal_2024}, or by modeling multi-$J$ CO and CO isotopologue lines to determine the underlying physical conditions in the gas \citep[e.g.,][]{Israel_2020,Teng_etal_2022,Teng_etal_2023,He_etal_2024}.
Numerical simulations have also provided important insights on this topic by implementing realistic treatments of CO chemistry and radiative transfer while accounting for observational resolution and sensitivity limits \citep[e.g.,][]{Glover_Clark_2012,Narayanan_etal_2012,Gong_etal_2020,HuCY_etal_2022}.

Thanks to these works, we have now established several key trends in the CO-to-H$_2$ conversion factor, $\alphaCO$, as functions of local gas properties.
For example, $\alphaCO$ anti-correlates strongly with gas-phase metallicity, likely due to more abundant ``CO-dark'' gas in lower metallicity environments \citep[e.g.,][]{Glover_MacLow_2011,Accurso_etal_2017,Gong_etal_2020,HuCY_etal_2022}. 
$\alphaCO$ also depends on gas temperature and optical depth, which affect the emissivity of the gas.
These variations appear to correlate with the CO line width \citep[e.g.,][]{Teng_etal_2024}, line intensity \citep[e.g.,][]{Narayanan_etal_2012,Gong_etal_2020}, and other environmental properties \citep[e.g., stellar or total mass surface density; see][]{Bolatto_etal_2013,Chiang_etal_2024}. 
One can therefore use these more accessible quantities as empirical proxies for CO emissivity variations when direct constraints are not available.
Besides, one may also need to account for CO excitation effects that alter the ratio between different CO transitions and thus their corresponding $\alphaCO$ in relation to each other \citep[e.g.,][]{Yajima_etal_2021,Leroy_etal_2022,denBrok_etal_2023,Keenan_etal_2024,denBrok_etal_2025}.
As with other factors affecting the CO emissivity, excitation variations appear to correlate with local conditions, especially the surface density of recent star formation.

These studies yield prescriptions that predict $\alphaCO$ from directly observable quantities, including metallicity, the surface densities of stellar mass and recent star formation rate (SFR), and the cloud-scale gas surface density or velocity dispersion. 
Obtaining estimates of these input variables requires ancillary data beyond just CO mapping, e.g., near-IR images tracing stellar mass, high-resolution millimeter line observations tracing gas kinematics, and/or metallicity estimates.
However, not all galaxies targeted for CO observations have such datasets readily available, and many modern CO studies still assume a constant $\alphaCO$ when converting CO emission to molecular gas mass. 
Unfortunately, this introduces systematic biases and can substantially affect the inferred relationships between gas mass and star formation \citep[\citealt{Bolatto_etal_2013}; also see][]{Sun_etal_2023,Teng_etal_2024,Leroy_etal_2025}.

This paper aims to address this problem by providing $\alphaCO$ predictions for thousands of local galaxies.
We follow current best practices \citep{Schinnerer_Leroy_2024} and apply the aforementioned empirical $\alphaCO$ calibrations homogeneously to the $z{=}0$ Multiwavelength Galaxy Synthesis \citep[$z$0MGS;][]{Leroy_etal_2019} dataset.
This atlas includes star-forming galaxies more massive than the LMC out to $d \approx 50$~Mpc, and so covers almost all targets in various local universe CO mapping campaigns, including COMING \citep{Sorai_etal_2019}, HERACLES \citep{Leroy_etal_2009}, the NRO Atlas \citep{Kuno_etal_2007}, PHANGS--ALMA \citep{Leroy_etal_2021a,Leroy_etal_2021b}, as well as VERTICO \citep{Brown_etal_2021} and its high-resolution successor MAUVE--ALMA (J.~Sun et al., in preparation). 
Together with the $\alphaCO$ predictions, we produce a variety of intermediate data products, including new estimates of galaxy effective radii and radial profiles of UV and IR band surface brightness, stellar mass surface density, and SFR surface density.
These intermediate calculations can also serve as a basis for more refined $\alphaCO$ estimates as $\alphaCO$ prescriptions and metallicity measurements improve in the future.

Beyond providing $\alphaCO$ estimates for various targets and surveys, this work also aims to offer a population-level view of $\alphaCO$ variations based on current prescriptions.
Specifically, we examine how $\alphaCO$ changes as functions of galactocentric radius, galaxy-integrated stellar mass, and star formation rate.
This allows us to predict how the ratio of star formation rate to CO luminosity, $\SFR/\LCO$, should vary at a given molecular gas depletion time ($\tdep \equiv \Mmol/\SFR$).
High values of $\SFR/\LCO$ in low-mass, low-metallicity galaxies have been observed for decades \citep[e.g.,][]{Young_Scoville_1991,Saintonge_etal_2011,Schruba_etal_2012,Genzel_etal_2012,Leroy_etal_2013,Hunt_etal_2020}, but the degree to which such variations reflect real changes in $\tdep$ or merely $\alphaCO$ is often unclear.
Our synthetic calculations provide a basis for interpreting such observations.

The structure of this paper is as follows.
Section~\ref{sec:method} describes the UV and IR datasets used in this work as well as the procedures for constructing surface brightness radial profiles, converting them into physical properties, and deriving $\alphaCO$ predictions based on empirical prescriptions.
Section~\ref{sec:compare} validates our $\alphaCO$ predictions against observational estimates for various subsets of galaxies in the literature.
Section~\ref{sec:trends} presents key population-level trends in the $\alphaCO$ predictions and examines their physical origins.
Section~\ref{sec:implications} discusses implications for galaxy-integrated molecular gas depletion times.
We summarize our main findings and describe the data products in Section~\ref{sec:summary}.


\section{Data \& Methods}
\label{sec:method}

\begin{figure*}[htb]
\gridline{
\fig{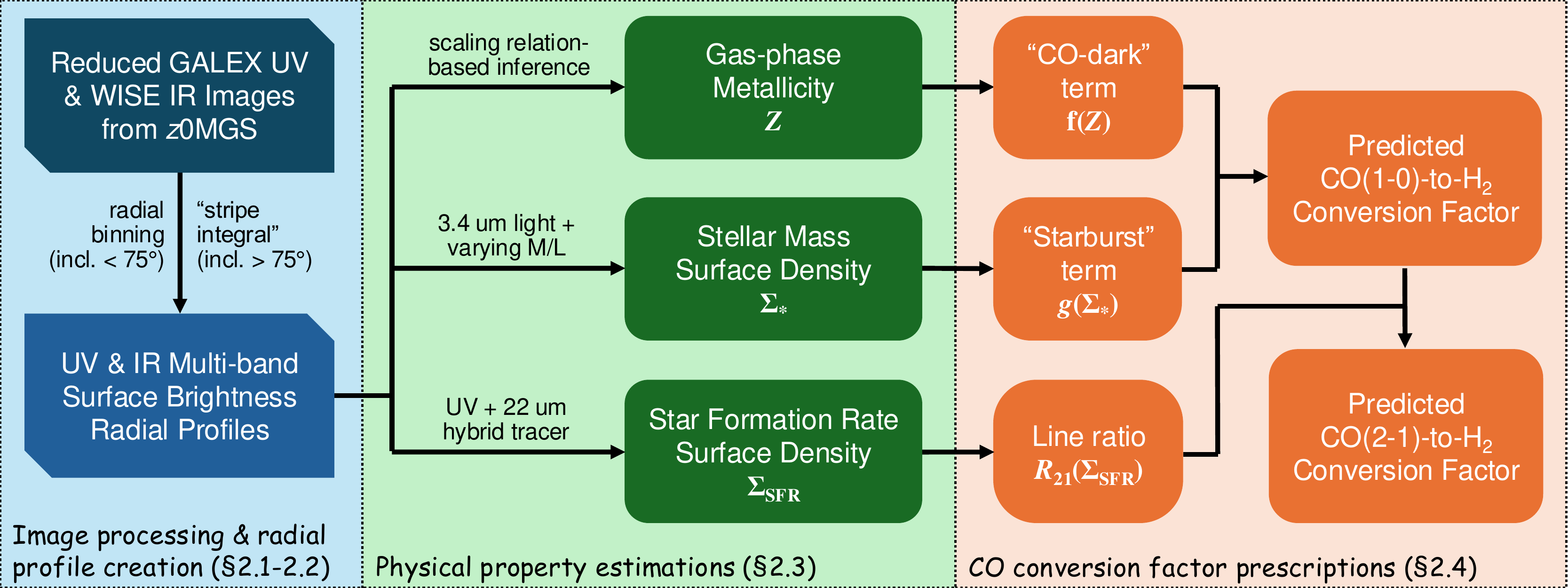}{\textwidth}{}
}
\vspace{-2\baselineskip}
\caption{
Our data processing workflow, from reduced GALEX and WISE images to $\alphaCO$ predictions.
Methodological details for the three major steps (image processing and radial profile creation, physical property estimation, and conversion factor prescription) are described in Sections~\ref{sec:method:data}-\ref{sec:method:radprof}, \ref{sec:method:phys}, and \ref{sec:method:alphaCO} respectively.
}
\vspace{0.5\baselineskip}
\label{fig:workflow}
\end{figure*}

We work with an extensive sample of local galaxies selected from the $z$0MGS sample \citep[][hereafter \citetalias{Leroy_etal_2019}]{Leroy_etal_2019}.
This parent sample includes over 15,000 galaxies in the HyperLEDA database \citep{Makarov_etal_2014}, most of which have \textit{B}-band absolute magnitudes $M_B\,{\lesssim}\,{-}18$~mag (i.e., comparable to or brighter than the LMC) and distances $d\,{\lesssim}\,50$~Mpc \citepalias[see figure~19 in][]{Leroy_etal_2019}.
Thanks to $z$0MGS, these galaxies have tabulated global properties as well as science-ready GALEX UV and WISE near-/mid-IR images, which allow us to estimate parameters such as stellar mass surface density and SFR surface density that are relevant to CO-to-H$_2$ conversion factor prescriptions.

From the $z$0MGS parent sample, we select galaxies for which it would be possible to estimate the \CO10-to-H$_2$ and \CO21-to-H$_2$ conversion factors following \citet[hereafter \citetalias{Schinnerer_Leroy_2024}]{Schinnerer_Leroy_2024} and for which we expect those prescriptions to apply. 
Specifically, we select galaxies with estimated global stellar mass $\Mstar\;{>}\;10^{9.3}\,\uM$ (i.e., comparable to or above LMC mass), similar to those galaxies used for calibrating the conversion factor prescriptions \citepalias[see references in][]{Schinnerer_Leroy_2024}.
We also require these galaxies to have known inclination and position angles because our analysis framework requires calculating galactocentric radius (see \autoref{sec:method:radprof} below).
We omit M31 and M33 partly due to challenges in processing their data given their enormous sky footprints, and partly because each already has their own conversion factor literature, which has in some cases informed the prescriptions we use \citep[e.g.,][]{Leroy_etal_2011,Smith_etal_2012,Gratier_etal_2017,Williams_etal_2019,Forbrich_etal_2020,Viaene_etal_2021}.
These selection criteria yield a total of \ngalall\ galaxies.

For each galaxy, we use UV images from GALEX \citep{Martin_etal_2005} and IR images from WISE \citep{Wright_etal_2010} to create UV and IR surface brightness radial profiles.
We then convert these observed quantities into physical properties and use them to predict the CO-to-H$_2$ conversion factors.
This data processing workflow is illustrated in \autoref{fig:workflow}, and the individual steps are detailed in the following subsections.

\subsection{UV and IR Images}
\label{sec:method:data}

We make use of GALEX FUV and NUV images (154 and 231~nm) as well as WISE1 through WISE4 band images (3.4, 4.6, 12, and 22~$\micron$) that were processed by the $z$0MGS project \citepalias{Leroy_etal_2019}.
These images were background-subtracted and convolved to Gaussian point spread functions (PSFs) of $7\farcs5$ (possible for all bands except WISE4 22~$\micron$) and $15''$ (possible for all bands).
The $15''$ resolution images reach typical 1$\sigma$ noise levels of ${\sim}1.5\times10^{-4}\;\uI$ in the GALEX bands, ${\sim}2.5\times10^{-3}\;\uI$ in the WISE1 3.4~$\micron$ band, and ${\sim}1.3\times10^{-1}\;\uI$ in the WISE4 22~$\micron$ band.

We also use star and galaxy masks published by \citetalias{Leroy_etal_2019} to mask pixels contaminated by bright foreground stars or other galaxies in the field.
However, we do not apply the masks close to galaxy centers because nuclear features in galaxies can sometimes be misidentified as stars and flagged in the masks.
Based on visual inspections of the images and masks, we set the overriding area to be within $0.15 \times R_{25}$ of the galaxy center or within one PSF from the galaxy center, whichever is larger.

For a subset of galaxies either located close to the Galactic plane or having close companion galaxies, a substantial fraction (${>}30\%$) of their UV/IR images can be masked, making it impractical to extract reliable measurements.
This issue is most severe in the WISE1 band (affecting $\sim$3,000 galaxies) but becomes much less concerning in the GALEX FUV and WISE4 bands (only $\sim$200 galaxies).
We omit these galaxies in the problematic bands in all following analyses (also see \autoref{tab:sample}).

\subsection{UV and IR Surface Brightness Profiles}
\label{sec:method:radprof}

\begin{figure*}[htb]
\gridline{
\fig{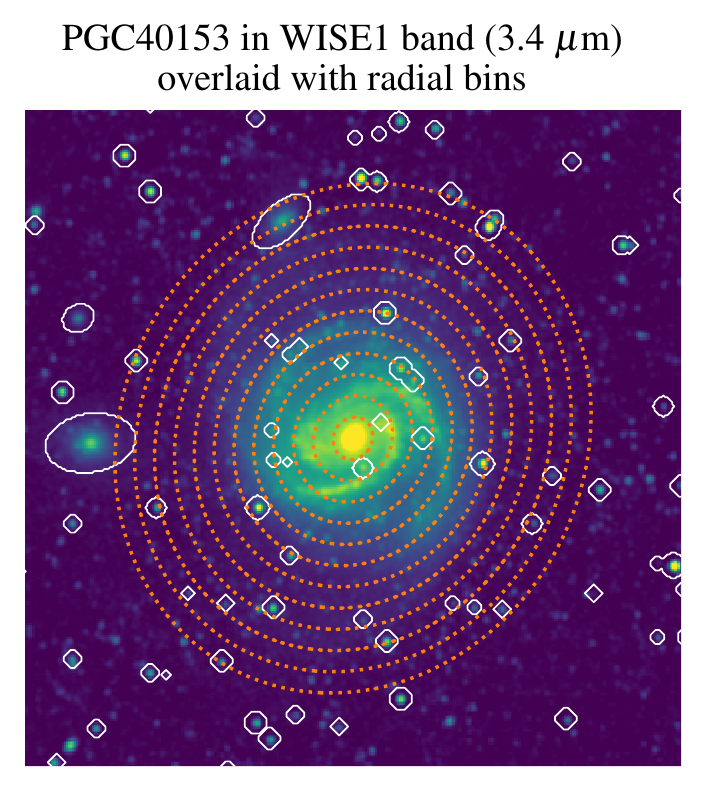}{0.33\textwidth}{}
\fig{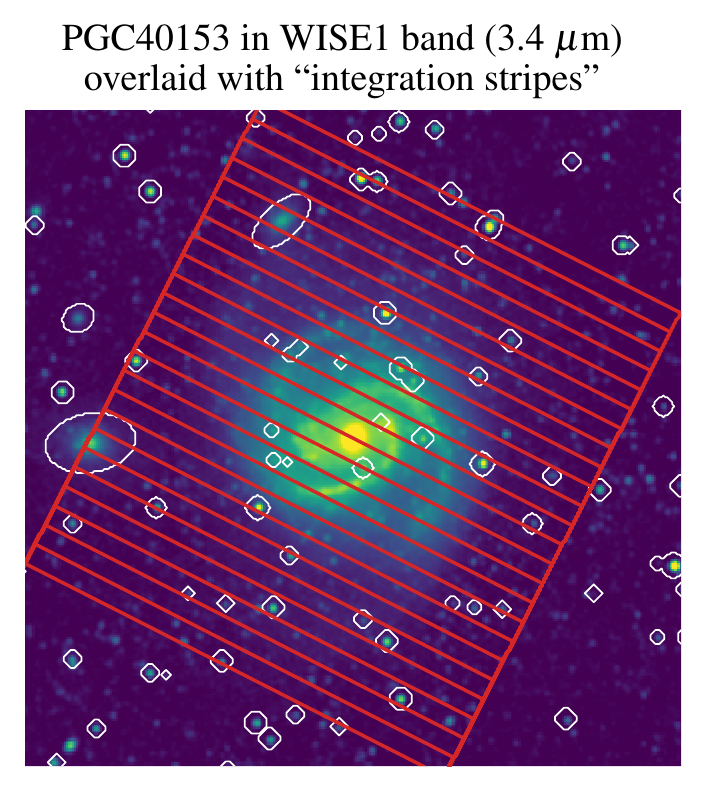}{0.33\textwidth}{}
\fig{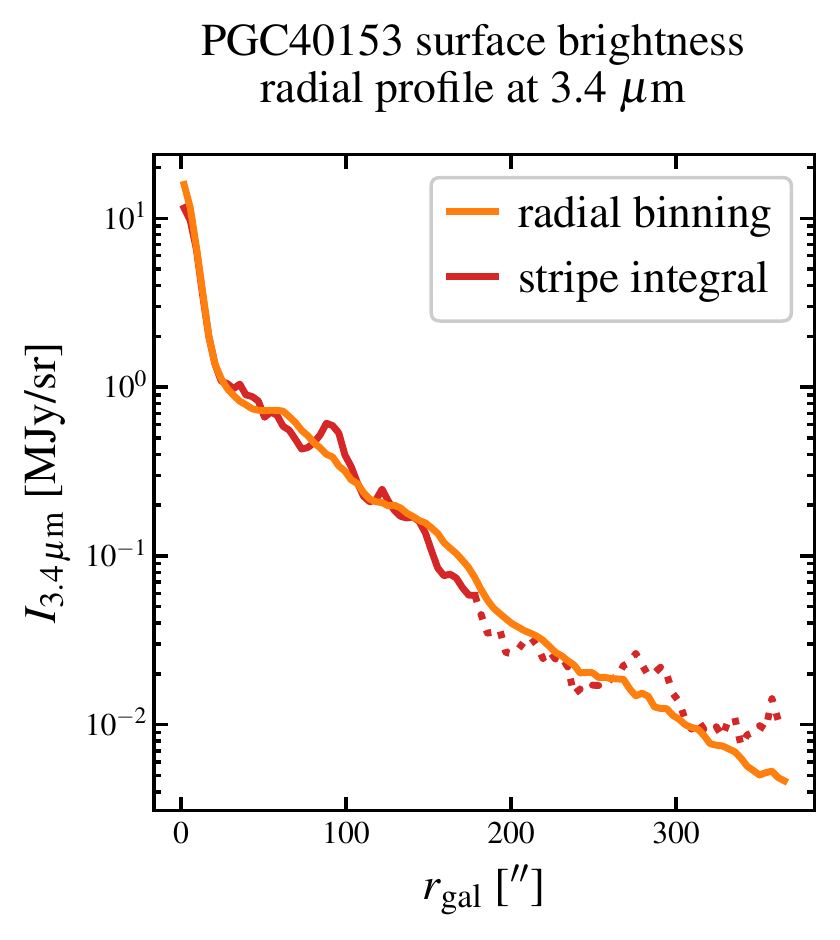}{0.329\textwidth}{}
}
\vspace{-2\baselineskip}
\caption{
The WISE1 band image (\textit{left \& middle panels}) and surface brightness radial profiles (\textit{right panel}) for PGC~40153 (a.k.a.\ NGC~4321, $i=27^\circ$), as an example of our radial profile construction techniques.
A set of elliptical rings (orange dotted lines) in the left panel represent galactocentric radius ($r\sbsc{gal}$) bins, in which we directly compute the mean/median surface brightness.
A set of long stripes (red rectangles) in the middle panel represent the regions used for ``stripe integral'' \citep{Warmels_1988}, an alternative method for deriving radial profiles that is applicable to even edge-on galaxies (see \autoref{sec:method:radprof}).
For visual clarity, the densities of bins/stripes are reduced by a factor of eight in both panels.
Both methods account for masked area due to foreground stars or other galaxies (white ellipses).
As shown by the right panel, the radial profiles derived with both methods agree well out to large $r\sbsc{gal}$, even after the stripe integral-based results drop below 3$\sigma$ significance (red dotted curve).
}
\vspace{0.5\baselineskip}
\label{fig:annuli_vs_stripes}
\end{figure*}

We build surface brightness radial profiles for each galaxy in each of the GALEX and WISE bands at both $7\farcs5$ and $15''$ resolution (which translates to ${\sim}1{-}3$~kpc physical scale for most galaxies).
For galaxies with inclination angle $i\leq75^\circ$, we build the radial profiles directly by generating a series of radial bins according to the inclination and position angle of the galaxy (see \autoref{fig:annuli_vs_stripes} left panel).
The radial bin width matches the half-width-half-maximum (HWHM) of the Gaussian PSF to appropriately sample the corresponding image.
The outermost bin reaches at least $2\,\radius{25}$ for each galaxy, and we require a minimum of 10 radial bins to have reasonable coverage even for the smallest galaxies.
We then calculate the mean surface brightness within each bin and the corresponding uncertainty via standard error propagation.
For WISE1 and WISE2 bands, we use the median surface brightness instead of the mean for all bins outside $0.5\,\radius{25}$ in order to suppress the residual effects of foreground stars \citep{Leroy_etal_2019}.
Finally, we multiply the surface brightness profiles by $\cos{i}$ to derive inclination-corrected surface brightness measurements and errors.
`
For galaxies with inclination $i>75^\circ$, direct radial binning is no longer reliable --- the exact inclination angle is challenging to measure in this regime, and the surface brightness profile becomes largely unresolved along the galactic minor axis in many cases.
In these cases, we instead use the ``stripe integral'' technique \citep{Warmels_1988} to reconstruct the radial surface brightness profiles.
In short, we calculate integrated flux densities within a series of ``integration stripes'' that align with the galactic minor axis and tile along the major axis (see \autoref{fig:annuli_vs_stripes} middle panel).
Assuming an axisymmetric disk and optically thin emission, we use this set of flux density measurements to derive the radial surface brightness profile via an inverse Abel transform.
This method is described in detail in Appendix~\ref{apdx:stripe}.

The stripe integral approach does not require knowledge of the galaxy inclination angle and can be applied to edge-on systems as long as all the assumptions hold.
As a check, we compare surface brightness profiles measured from direct radial binning to those inferred with the stripe integral method for relatively face-on galaxies (see \autoref{fig:annuli_vs_stripes} right panel for an example).
We find that the two approaches yield consistent results for the vast majority of targets -- among those with high-quality profiles derived from both methods, over 80\% of targets have their bin-to-bin surface brightness values agree within $30\%$ (or $\lesssim0.1$~dex), and over 90\% show agreements within $50\%$ ($\lesssim0.2$~dex).
Inconsistent radial profiles typically occur in cases where: (a) the cataloged galaxy inclination and/or position angle are potentially wrong, (b) the surface brightness profile is only marginally resolved in the given band, or (c) the signal-to-noise (S/N) ratio of the detection becomes low, especially far into the galaxy outskirts.
While there is no easy solution in the first case, in the latter two cases we either flag the galaxy (when it is unresolved) or the radial bins with low S/N ratio to avoid unreliable measurements in subsequent analyses (see \autoref{sec:method:radprof:r50} below).

\begin{deluxetable}{lc}
\tablecaption{Number of Galaxies with Key Measurements\label{tab:sample}}
\tablewidth{0pt}
\tablehead{
\colhead{Sample Description} & 
\colhead{$N\sbsc{face\text{-}on}/N\sbsc{total}$} \\[-1.5em]
}
\startdata
\\[-1.5em]
$z$0MGS parent sample \citep{Leroy_etal_2019} & ${-}/15,748$ \\
Full sample considered in this work (\S\ref{sec:method}) & $6,868/10,657$ \\
\hline
\multicolumn{2}{c}{Galaxies with images \& surface brightness profiles (\S\ref{sec:method:data}--\ref{sec:method:radprof})} \\
\hline
... for GALEX FUV ($154\,\text{nm}$) & $4,826/7,554$ \\
... for GALEX NUV ($231\,\text{nm}$) & $4,747/7,415$ \\
... for WISE1 ($3.4\,\micron$) & $4,897/7,613$ \\
... for WISE2 ($4.6\,\micron$) & $5,175/8,053$ \\
... for WISE3 ($12\,\micron$) & $6,290/9,766$ \\
... for WISE4 ($22\,\micron$) & $6,708/10,416$ \\
\hline
\multicolumn{2}{c}{Galaxies with physical measurements or predictions} \\
\hline
WISE1 half-light radius (\S\ref{sec:method:radprof:r50}) & $4,848/7,279$ \\
Stellar \& SFR surface density profiles (\S\ref{sec:method:phys}) & $3,956/5,300$ \\
CO-to-H$_2$ conversion factor profile (\S\ref{sec:method:alphaCO}) & $3,919/5,244$ \\
\enddata
\tablecomments{
$N\sbsc{face\text{-}on}$ counts galaxies with inclination $i\leq75^\circ$, for which we have measurements based on direct radial binning; $N\sbsc{total}$ counts all galaxies with either direct radial binning or stripe integral-based measurements (see \S\ref{sec:method:radprof}).
}
\vspace{-2\baselineskip}
\end{deluxetable}

\subsubsection{Galaxy Half-light Radii}
\label{sec:method:radprof:r50}

From the surface brightness radial profiles measured with either direct radial binning or stripe integral, we derive the half-light radius (or effective radius, $\radius{50}$) of each galaxy in each UV/IR band.
This involves calculating the cumulative flux distribution as a function of galactocentric radius and determining the radius at which it reaches 50\% of the galaxy total flux.
We use the total fluxes reported in \citetalias{Leroy_etal_2019} and note that the numbers vary by only $\sim$0.02~dex with mildly different methodological choices (such as the area for overriding star masks or calculating the cumulative flux, see \autoref{sec:method:data}--\ref{sec:method:radprof}).

For galaxies with sizes comparable to or smaller than the image resolution, the calculated $\radius{50}$ can be biased high because the galaxy radial profiles are not well resolved.
To address this issue, we generate mock galaxies with varying sizes and inclination angles, convolve their images to the typical data resolution we work with, measure $\radius{50}$ with the method described above, and compare those measurements with the ground truth values (see \autoref{apdx:res} for more details).
This exercise produces a correction factor that depends on the ratio of the measured radius to the PSF size, $\radius{50,\,obs}/\theta_\mathrm{ PSF}$, and for radial profiles (but not stripe integral-based sizes) the galaxy inclination angle.
We calculate this factor for each profile and scale $\radius{50,\,obs}$ accordingly.
This yields our best estimate $\radius{50}$ corrected for the effects of resolution.

We flag a small subset of galaxies whose $\radius{50}$ measurements are deemed unreliable for one of the following reasons: (1) the surface brightness profile drops below 3$\sigma$ significance (per bin) before reaching $\radius{50}$, which leads to large error on the $\radius{50}$ measurement; (2) the inferred $\radius{50}$ lies within the first radial bin, which means their surface brightness profile is completely unresolved; or (3) the galaxy is marginally resolved, but the estimated resolution bias correction on $\radius{50}$ is larger than a factor of 2, in which case the error on $\radius{50}$ would be too large after correction (see \autoref{apdx:res}).
The fraction of flagged galaxies varies across different bands and differs between the radial binning-based sizes and the stripe integral-based ones.
For the WISE1 band at $7\farcs5$ resolution, in total $<10\%$ of galaxies with stripe integral-based profiles are flagged, and $<1\%$ of those with radial binning-based profiles are flagged (see \autoref{tab:sample}).

\subsection{From Observed Quantities to Physical Properties}
\label{sec:method:phys}

Based on the measured surface brightness radial profiles and half-light radii for each galaxy, we calculate the star formation rate (SFR) surface density, stellar mass surface density, and gas-phase metallicity, all of which are input parameters for the \citetalias{Schinnerer_Leroy_2024} conversion factor prescription.
We summarize the key steps below and refer interested readers to \citetalias{Leroy_etal_2019} and \citet{Sun_etal_2022} for detailed descriptions.

We derive SFR surface densities ($\SigSFR$) from GALEX UV and WISE mid-IR surface brightnesses closely following \citetalias{Leroy_etal_2019} (see Table~7 and appendix therein).
By default, we combine GALEX FUV (154~nm) with WISE4 (22~$\micron$) to trace both exposed and obscured star formation.
For targets without GALEX FUV data, we combine GALEX NUV (231~nm) with WISE4 whenever NUV is available.
We resort to a WISE4-only calibration when neither FUV nor NUV data is available.
Due to the use of WISE4 data, our $\SigSFR$ results are only estimated at $15''$ resolution.

We derive stellar mass surface densities ($\Sigstar$) based on WISE1 3.4~$\micron$ surface brightness profiles ($I\sbsc{3.4\mu m}$) at $7\farcs5$ resolution and a radially varying stellar mass-to-light (M/L) ratio.
We estimate the latter from the local $\SigSFR$-to-$I\sbsc{3.4\mu m}$ ratio, following an empirical calibration presented in \citetalias{Leroy_etal_2019} (Table~6 therein).
This method accounts for M/L trends associated with varying stellar population ages across galaxies.

We further apply a $\text{S/N}>3$ threshold (per radial bin) on the derived $\SigSFR$ and $\Sigstar$ radial profiles, so that low-significance measurements are considered non-detections instead.
Given this threshold, a fraction of galaxies do not have significant $\SigSFR$ or $\Sigstar$ measurements in any radial bins.
This is mainly due to the limited sensitivity of the WISE4 images, which are directly used for $\SigSFR$ and indirectly affects the M/L ratio and therefore $\Sigstar$.
Among the galaxies with measured surface brightness profiles in all relevant GALEX and WISE bands, \ngalstarsfr\ galaxies have significant $\SigSFR$ and $\Sigstar$ measurements;
\ngalstarsfrrp\ out of these \ngalstarsfr\ targets have low or moderate inclination ($i\,{\leq}\,75^\circ$) and therefore have direct radial binning measurements (see \autoref{sec:method:radprof}).

For gas-phase metallicity ($Z$), since direct measurements are not available for the majority of galaxies in our sample, we rely on empirical scaling relations to estimate metallicity following \citet{Sun_etal_2020b,Sun_etal_2022}.
Specifically, we use the galaxy global stellar mass published in \citetalias{Leroy_etal_2019} to estimate the metallicity at $r\,{=}\,r_{50}$ in each galaxy, following the mass-metallcity relation measured by the SAMI survey \citep{Sanchez_etal_2019}.
We then use our measured $r_{50}$ in WISE1 band from \autoref{sec:method:radprof} and a radial metallicity gradient of $-0.1\,\mathrm{dex}/r_{50}$ \citep[from CALIFA;][]{Sanchez_etal_2014} to derive the implied metallicity elsewhere in the galaxy.
For these calculations, we adopt the O3N2 metallicity calibration \citep{Pettini_Pagel_2004} and a solar value of $12+\log\mathrm{(O/H)}=8.69$~dex \citep{Asplund_etal_2009}.

This approach allows us to obtain uniform metallicity estimates across the full sample.
Considering that the $z$0MGS parent sample covers similar galaxy populations as the SAMI and CALIFA surveys (in terms of stellar mass range etc.), our metallicity estimates should statistically match the average trends seen in IFU-based metallicity measurements across many galaxies.
That being said, there are appreciable galaxy-to-galaxy variations in both the mass-metallicity relation and the radial metallicity gradient \citep[e.g., see][]{Kreckel_etal_2019,Sanchez_2020}, so that we do not necessarily expect our scaling relation-based estimates to perfectly match the observed trends in individual galaxies.
For studies that focus on smaller sets of galaxies with uniform optical IFU coverage \citep[e.g., EDGE-CALIFA, PHANGS, MAUVE, KILOGAS;][]{Sanchez_etal_2013,Kreckel_etal_2019,Catinella_etal_2025}, it is possible to improve upon our approach by directly incorporating emission line-based metallicity measurements.

\subsection{Conversion Factor Prescriptions}
\label{sec:method:alphaCO}

We predict CO-to-H$_2$ conversion factors for both \CO10\ and \CO21\ lines with a prescription recommended by \citetalias{Schinnerer_Leroy_2024}.
This prescription involves three terms that account for the effects of CO-dark gas, CO emissivity variations, and CO excitation effects, respectively.
The ``CO-dark'' term is parameterized as a function of the gas-phase metallicity $Z$:
\begin{align}
&f(Z) = \left(Z/Z_\odot\right)^{-1.5} \label{eq:CO-dark}\\
(&\text{for}~0.2 < Z/Z_\odot < 2). \nonumber
\end{align}
\noindent Here, the power-law slope of $-1.5$ is recommended by \citetalias{Schinnerer_Leroy_2024} and broadly consistent with various observational constraints from \CII, dust, and gas depletion time measurements \citep[e.g.,][]{Schruba_etal_2012,Amorin_etal_2016,Accurso_etal_2017,Hunt_etal_2020}.
We also calculate an alternative $f(Z)_\mathrm{G20}$ with a shallower power law index of $-0.8$ \citep[following][]{Gong_etal_2020} for comparison, as numerical simulation studies often predict relatively shallow slopes \citep[also see][]{Glover_Clark_2012,HuCY_etal_2022}.
Considering that \autoref{eq:CO-dark} is mostly calibrated for $Z \sim 0.2{-}2.0\,Z_\odot$, we limit the power-law dependence to within this metallicity range and use the boundary values outside this range\footnote{Note that metallicity values beyond $0.2{-}2.0\,Z_\odot$ are very rare in our sample, partly due to the sample selection scheme (especially $\Mstar>10^{9.3}\,\uM$; see \autoref{sec:method}).} (i.e., use $f(2Z_\odot)$ at $2.5\,Z_\odot$).

The ``emissivity'' term aims at capturing variations in CO optical depth and/or excitation temperature.
Following \citet{Bolatto_etal_2013} and \citet{Chiang_etal_2024}, it is parameterized as a function of the local stellar mass surface density $\Sigstar$:
\begin{align}
&g(\Sigstar) = \left(\frac{\Sigstar}{100\,\uSig}\right)^{-0.25} \label{eq:emissivity}\\
&(\text{for}~\Sigstar > 100\,\uSig). \nonumber
\end{align}
\noindent The power-law index of $-0.25$ originates from \citet{Chiang_etal_2024} based on \CO10\ data, though a steeper $-0.5$ slope was suggested by \citet{Bolatto_etal_2013} based on mostly \CO21\ data.
We calculate both in this work for comparisons.
In either case, this $g(\Sigstar)$ term is only in effect at $\Sigstar > 100\,\uSig$, where substantial changes in CO emissivity tend to occur based on empirical data \citep{Bolatto_etal_2013,Chiang_etal_2024}.

The CO line ratio $R_{21} \equiv \ICOxy21/\ICOxy10$ also varies with excitation condition. \citetalias{Schinnerer_Leroy_2024} suggested a parameterization involving the local SFR surface density $\SigSFR$:
\begin{align}
R_{21}(\SigSFR) &= 0.65 \left(\frac{\SigSFR}{0.018\,\uSigSFR}\right)^{0.125} \label{eq:R21}\\
&(\text{for}~0.35 < R_{21} < 1.0). \nonumber
\end{align}
\noindent The $0.125$ power law index recommended by \citetalias{Schinnerer_Leroy_2024} is updated from \citet{Leroy_etal_2022} and broadly agrees with other studies in the literature \citep[e.g.,][]{denBrok_etal_2021,Yajima_etal_2021,denBrok_etal_2025}.
We limit the predicted $R_{21}(\SigSFR)$ value to be within $[0.35, 1.0]$ (i.e., replace all values outside this range with the boundary values), following \citetalias{Schinnerer_Leroy_2024} and reflecting the range of $R_{21}$ values seen in most observations.

We combine all three terms described above to derive the conversion factors for both \CO10\ and \CO21:
\begin{align}
\alphaCOxy10
&= \alphaCOxy10^\mathrm{MW}\,f(Z)\,g(\Sigstar)~, \label{eq:alphaCO10}\\
\alphaCOxy21
&= \alphaCOxy10^\mathrm{MW}\,f(Z)\,g(\Sigstar)\,R_{21}(\SigSFR)^{-1}~,
\label{eq:alphaCO21}
\end{align}
\noindent where $\alphaCOxy10^\mathrm{MW}=4.35\,\ualphaCO$ is the commonly adopted Galactic value.
We feed the $Z$, $\Sigstar$ and $\SigSFR$ profiles calculated in \autoref{sec:method:phys} into \autoref{eq:alphaCO10}--\ref{eq:alphaCO21} to derive both $\alphaCOxy10$ and $\alphaCOxy21$ as functions of galactocentric radius in each galaxy.

We emphasize that these $\alphaCO$ predictions should be viewed to have effective resolution of $\gtrsim$kpc scales.
This is not only because the input GALEX and WISE images are at these resolution (see \autoref{sec:method:data}), but also because the \citetalias{Schinnerer_Leroy_2024} $\alphaCO$ prescriptions were largely motivated by and calibrated against kpc-scale measurements (see references therein).
Although we do expect $\alphaCO$ to vary on smaller, $\lesssim$100~pc scales and have seen evidence for that in simulations and observations \citep[e.g.,][]{Gong_etal_2018,Teng_etal_2022,Teng_etal_2023}, our $\alphaCO$ predictions are not designed to capture such variations, but rather to reflect (flux-weighted) average values over $\gtrsim$kpc area.


\section{Comparisons to Conversion Factor Measurements in the Literature}
\label{sec:compare}

To verify the reliability of our $\alphaCO$ predictions, we compile observational measurements of $\alphaCO$ from the literature \citep{Sandstrom_etal_2013,Israel_2020,Teng_etal_2022,Teng_etal_2023,denBrok_etal_2023,Yasuda_etal_2023,Chiang_etal_2024}.
Some of these works even helped motivate and calibrate the \citetalias{Schinnerer_Leroy_2024} prescription in the first place.
They all focus on smaller sets of nearby galaxies within our sample, making it possible to directly benchmark our $\alphaCO$ predictions using these galaxies.
They also rely on different types of observations and employ different methods to measure $\alphaCO$, as detailed below:

\begin{itemize}[itemsep=0.5em,leftmargin=1em,parsep=0em,partopsep=0em]

\item[$\sbullet$] \emph{\citet[dust-based]{Sandstrom_etal_2013}:} This work combines \CO21, \HI, and far-IR dust observations to simultaneously solve for $\alphaCO$ and the dust-to-gas (D/G) ratio in 26 galaxies. These measurements have an effective resolution of $75''$ (i.e., size of ``solution pixels'' therein). Solely based on \CO21\ data, the $\alphaCO$ measurements are essentially $\alphaCOxy21$, although they were expressed as $\alphaCOxy10$ assuming a fixed $R\sbsc{21}$.

\item[$\sbullet$] \emph{\citet[dust-based]{denBrok_etal_2023}:} This work combines CO, \HI, and dust data to solve for $\alphaCO$ and D/G ratio in two galaxies (M51 and M101). The method closely follows \citet{Sandstrom_etal_2013}, and the effective resolution is also $75''$. Both $\alphaCOxy10$ and $\alphaCOxy21$ are determined in this work as it incorporates both \CO10\ and \CO21\ data.

\item[$\sbullet$] \emph{\citet[dust-based]{Yasuda_etal_2023}}: This work combines \CO10, \HI, and dust data to solve for $\alphaCO$ and D/G ratio for 22 galaxies. The method largely follows \citet{Sandstrom_etal_2013}, except that it measures one $\alphaCO$ value per galaxy, within an effective area that varies from galaxy to galaxy. Only $\alphaCOxy10$ is available as no \CO21\ data were used.

\item[$\sbullet$] \emph{\citet[dust-based]{Chiang_etal_2024}:} This work combines CO, \HI, dust, and metallicity measurements to derive $\alphaCO$ for 25 galaxies with \CO10\ and 28 galaxies with \CO21. The effective resolution of these measurements are 2~kpc in physical scale, or $20''{-}200''$ in angular size depending on the distance to each galaxy. $\alphaCOxy10$ and $\alphaCOxy21$ are measured for galaxies with \CO10\ and \CO21\ data, respectively.

\item[$\sbullet$] \emph{\citet[carbon budget accounting]{Israel_2020}}: This work compiles multi-$J$ CO, [\CI], and [\CII] line data to account for the overall carbon budget and derive $\alphaCO$ for 69 galaxies. These measurements are for the central ${\sim}20''$ in each galaxy. With $R\sbsc{21}$ measured from observations, this work effectively determines both $\alphaCOxy10$ and $\alphaCOxy21$ at the same time.

\item[$\sbullet$] \emph{\citet[LVG modeling]{Teng_etal_2022,Teng_etal_2023}:} These works model multi-$J$ CO and CO isotopologue lines to constrain the gas physical conditions and $\alphaCO$ in the center of three galaxies (NGC~3351, 3627, and 4321). Although the original measurements have high native resolution (${\sim}1''$), we use flux-weighted $\alphaCO$ values across the entire central ${\sim}30''$ in these galaxies for comparisons with our $\alphaCO$ predictions. Both $\alphaCOxy10$ and $\alphaCOxy21$ are available from these works.

\end{itemize}

We compare these $\alphaCO$ measurements with our predictions \textit{at matched effective resolution}.
That is, for each target from each literature work, we ``rebin'' its predicted $\alphaCO$ radial profiles by merging every $N$ radial bins into one wider bin, such that the new bin width matches the effective resolution of the observational measurements.
We calculate a weighted\footnote{We use the WISE3 $12\,\micron$ flux \citep[which strongly correlates with CO; see][]{Chown_etal_2021} as the weight for each bin, so that each bin is approximately weighted by the expected total CO flux it encloses.} average $\alphaCO$ value across the contributing radial bins to determine the appropriate $\alphaCO$ prediction for each new, wider bin.
These ``rebinned'' predictions are then compared with the median value of observational $\alphaCO$ measurements falling inside each (new) radial bin.

\subsection{\texorpdfstring{CO\,(1--0)}{CO(1-0)} Conversion Factor}
\label{sec:compare:alphaCO10}

\begin{figure*}[p]
\gridline{
\fig{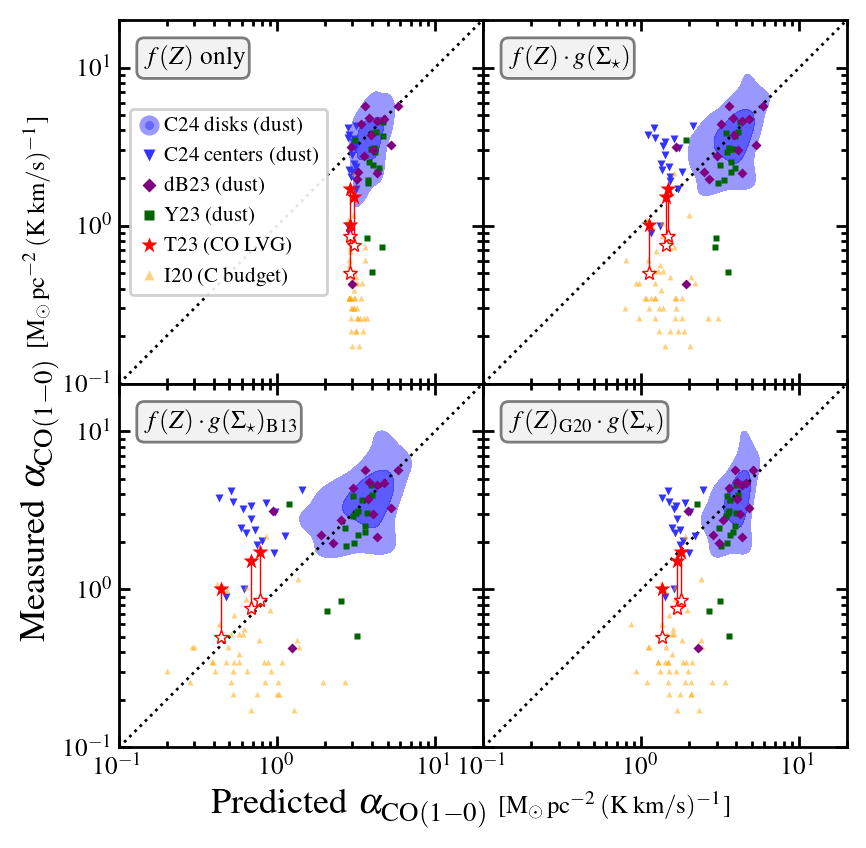}{0.7\textwidth}{}
}
\vspace{-2.5\baselineskip}
\caption{
Comparing four versions of \CO10-to-H$_2$ conversion factor predictions ($x$-axes) with observational estimates from the literature ($y$-axis).
The latter include dust-based estimates across many galaxies presented in \citet[dB23]{denBrok_etal_2023}, \citet[Y23]{Yasuda_etal_2023}, and \citet[C24]{Chiang_etal_2024}, as well as CO multi-line modeling (i.e., LVG) by \citet[T23]{Teng_etal_2023} and carbon budget accounting by \citet[I20]{Israel_2020} in galaxy centers.
Note that the blue contours and downward triangles separately show measurements in galaxy disks and centers from \citet{Chiang_etal_2024}; the filled and open stars show two versions of results from \citet{Teng_etal_2023} assuming CO/\Htwo\ abundance ratios of $1.5{\times}10^{-4}$ and $3{\times}10^{-4}$, respectively.
\textit{Top left:} The metallicity-dependent CO-dark term $f(Z)$ alone only spans a limited dynamic range, thus failing to match low $\alphaCOxy10$ values from observational estimates in galaxy centers.
\textit{Top right:} A combination of the $f(Z)$ term and a $\Sigstar$-dependent emissivity term $g(\Sigstar)$ matches dust-based and LVG-based estimates well but still disagrees with carbon budget accounting results \citep{Israel_2020}.
\textit{Bottom left:} Adopting an alternative emissivity term $g(\Sigstar)\sbsc{B13}$ \citep[][]{Bolatto_etal_2013} results in lower $\alphaCOxy10$ than most estimates for galaxy centers though still moderately higher than those from \citet{Israel_2020}.
\textit{Bottom right:} An alternative CO-dark term $f(Z)\sbsc{G20}$ \citep{Gong_etal_2020} yields similar results as the fiducial choice (top right), though the weaker metallicity dependence in $f(Z)\sbsc{G20}$ leads to slightly more underestimated $\alphaCO$ at the high end (i.e., in outer galaxy disks).
}
\vspace{0.3\baselineskip}
\label{fig:alphaCO10}
\end{figure*}

\autoref{fig:alphaCO10} shows these comparisons for $\alphaCOxy10$.
To assess the effect of each term in \autoref{eq:alphaCO10}, we show four versions of $\alphaCOxy10$ predictions.
The first shows only the metallicity term.
The other three show both metallicity and emissivity terms, but vary each term between our fiducial prescription and alternative, commonly adopted literature prescriptions.

The version without an emissivity term, i.e., $\alphaCOxy10=\alphaCOxy10\spsc{MW}\,f(Z)$, matches the observations reasonably well at the high $\alphaCO$ end, which mostly correspond to intermediate to large galactocentric radii.
However, this version of the $\alphaCOxy10$ prediction exhibits a limited dynamic range of $3{-}7\,\ualphaCO$ and is not able to capture the wide range of measured values, which span ${\sim}0.2{-}10\,\ualphaCO$.
This mismatch between predicted and estimated $\alphaCOxy10$ is particularly pronounced at low galactocentric radii (i.e., near galaxy centers) where the observed $\alphaCOxy10$ tends to be much lower than the $f(Z)$-only prediction, as highlighted by the \citet{Israel_2020} and \citet{Teng_etal_2023} measurements.

Including an emissivity term in the $\alphaCO$ predictions leads to much better agreement with the low $\alphaCOxy10$ values observed in galaxy centers.
With the fiducial functional form suggested by \citetalias{Schinnerer_Leroy_2024} (i.e., \autoref{eq:emissivity}), the predicted $\alphaCOxy10$ agrees well with most dust-based estimates over a wider dynamic range (see \autoref{fig:alphaCO10} top right); it also matches the LVG-based estimates when adopting a CO/\Htwo\ abundance ratio of $1.5{\times}10^{-4}$ following \citet[see filled stars in \autoref{fig:alphaCO10}]{Teng_etal_2024}.
However, the predicted values are still systematically higher than $\alphaCOxy10$ estimates by \citet{Israel_2020} based on carbon budget accounting\footnote{It is worth noting that the majority of $\alphaCO$ estimates from \citet{Israel_2020} are around the optically thin CO limit of $\alphaCOxy10\approx0.4\;\ualphaCO$ \citep[assuming a fiducial CO abundance of $1.5{\times}10^{-4}$ and an excitation temperature of 50~K]{Bolatto_etal_2013}, which means they are close to the lowest possible values.}.
This reflects an apparent tension between  \citet{Israel_2020} and dust-based studies like \citet{Chiang_etal_2024}, especially since the galaxy samples examined by these works overlap substantially.

Given this unresolved tension in the literature, we experiment with an alternative version of the emissivity term, with a steeper power law slope ($-0.5$) suggested by \citet{Bolatto_etal_2013}.
As \autoref{fig:alphaCO10} bottom left panel shows, this alternative prescription yields $\alphaCOxy10$ predictions that are substantially lower than most dust-based estimates in galaxy centers.
It may be consistent with LVG-based estimates provided that a higher CO/\Htwo\ abundance ratio of $3{\times}10^{-4}$ is assumed.
Nevertheless, the predictions are still higher than the \citet{Israel_2020} estimates.
To make them align, an even steeper slope would be required for the emissivity term, and the discrepancies with the dust-based and LVG-based results would become more severe \citep[also see][]{Downes_Solomon_1998,Dunne_etal_2022}.

We also experiment with an alternative CO-dark term with a shallower power-law index, as suggested by \citet{Gong_etal_2020}.
It appears to yield similar results as the fiducial functional form (\autoref{fig:alphaCO10} top right versus bottom right): for example, the ``disk'' measurements from \citet{Chiang_etal_2024} show ${\sim}0.2$~dex residual scatter around the identity line in both cases.
Overall, the current set of $\alphaCO$ measurements does not provide strong differentiating power between these CO-dark corrections.

\begin{figure*}[bt]
\gridline{
\fig{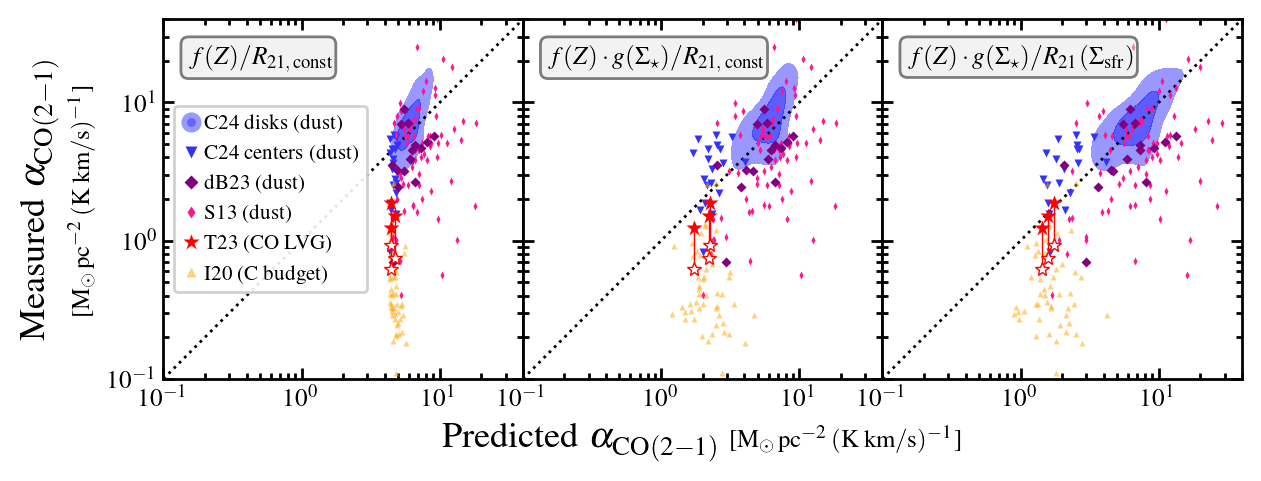}{\textwidth}{}}
\vspace{-2.5\baselineskip}
\caption{
Comparing three versions of \CO21-to-H$_2$ conversion factor predictions ($x$-axes) with observational estimates from the literature ($y$-axis).
The latter include a similar set of literature results as \autoref{fig:alphaCO10}, except omitting $\alphaCOxy10$ estimates by \citet{Yasuda_etal_2023} and adding $\alphaCOxy21$ from \citet[S13]{Sandstrom_etal_2013}.
\textit{Left \& middle:} Assuming a constant line ratio of $R_\mathrm{21,\,const}=0.65$, the CO-dark and emissivity terms in combination can provide a wider range of $\alphaCOxy21$ values and better agreements with observational results than the CO-dark term alone.
\textit{Right:} Including a $\SigSFR$-dependent line ratio term on top of the CO-dark and emissivity terms further improves the agreement with observations at both low and high $\alphaCOxy21$ ends.
}
\vspace{0.3\baselineskip}
\label{fig:alphaCO21}
\end{figure*}

\subsection{\texorpdfstring{CO\,(2--1)}{CO(2-1)} Conversion Factor}
\label{sec:compare:alphaCO21}

\autoref{fig:alphaCO21} shows predictions of $\alphaCOxy21$ in comparison to observational estimates from the literature.
The set of literature results included here are similar to \autoref{fig:alphaCO10}, except that we omit \citet[since this work did not measure $\alphaCOxy21$]{Yasuda_etal_2023} and include additional dust-based $\alphaCOxy21$ estimates by \citet{Sandstrom_etal_2013}.

Similar to the $\alphaCOxy10$ case, the observational estimates for $\alphaCOxy21$ span a wide dynamic range of ${\sim}0.2{-}20\,\ualphaCO$.
This is not captured by $\alphaCOxy21$ predictions that account for only the metallicity-dependent CO-dark term (i.e., omitting the emissivity term and adopting $R_\mathrm{21,\,const}=0.65$; \autoref{fig:alphaCO21} left panel).
Even though this approach gives a reasonable normalization for most measurements outside galaxy centers (e.g., going through the middle of the ``disk'' measurements from \citealt{Chiang_etal_2024}), it tends to under-predict measurements in galaxy outskirts (i.e., the high $\alphaCOxy21$ end) and severely over-predict in galaxy centers (the low $\alphaCOxy21$ end).

Including the fiducial emissivity term brings the predictions into better agreement with most observations in galaxy centers (\autoref{fig:alphaCO21} middle panel).
Nonetheless, the aforementioned tension between the dust-based method and carbon budget accounting is also present for $\alphaCOxy21$, such that the predictions cannot match both at the same time.
Besides, when incorporating both the CO-dark and emissivity terms but assuming a fixed line ratio of $R_\mathrm{21,\,const}=0.65$, the predictions are still lower than observational estimates at the high $\alphaCOxy21$ end.

The full prescription with all three terms (i.e., \autoref{eq:alphaCO21}) shows the best agreement with most observational estimates (\autoref{fig:alphaCO21} right panel).
The varying $R_{21}$ term is needed for achieving this agreement at the high $\alphaCOxy21$ end as $R_{21}$ often decreases systematically towards galaxy outskirts as $\alphaCO$ increases \citep[e.g.,][]{denBrok_etal_2021,Yajima_etal_2021,Leroy_etal_2022}.
This is supported by the reduced scatter around the identity line when including the $R_{21}$ term \citep[e.g., from 0.19 to 0.17~dex for the ``disk'' measurements from][]{Chiang_etal_2024}.
The variable $R_{21}$ term also helps in the case of galaxy centers, where $R_{21}$ can sometimes approach its thermal value (${\sim}1$).
The general effects of $R_{21}$ variations on $\alphaCOxy21$ are relatively mild, as $R_{21}$ has a very narrow dynamic range (${\sim}0.3{-}1$); these effects can nevertheless become evident over a sizable sample of galaxies \citep[also see \autoref{sec:trends:resolved}--\ref{sec:trends:global} and][]{Keenan_etal_2024}.

\subsection{Path to Better Conversion Factor Recipes}
\label{sec:compare:future}

The comparisons between $\alphaCO$ prescriptions and measurements in \autoref{sec:compare:alphaCO10}--\ref{sec:compare:alphaCO21} highlight several aspects where future improvements are needed.
For example, the CO-dark gas term is critical to understanding $\alphaCO$ variations in galaxies over a wide stellar mass range and as a function of radius within galaxies (also see \autoref{sec:trends} below).
A recurring issue related to this term is the challenge of obtaining precise and preferably direct metallicity estimates for systems where $\alphaCO$ estimates are available or required \citep[see discussion in][]{Chiang_etal_2024}.
In this sense, the emerging synergies between CO and optical IFU surveys (e.g., EDGE, PHANGS, MAUVE, KILOGAS) will enable a major step forward.
The CO-dark gas term is also expected to depend on gas surface density and dust-to-gas ratio in addition to metallicity \citep[e.g., see][]{Bolatto_etal_2013}.
Here, highly resolved estimates of $\alphaCO$ across diverse sub-galactic environments in local galaxies will help uncover dependence on secondary parameters.

More generally, we need a more extensive set of high-quality $\alphaCO$ measurements that span a wide dynamic range in metallicity and galaxy type.
This is essential for improving observational constraints on the CO-dark term and narrowing down from the wide range of calibrations present in the literature \citep[e.g., see][]{Schinnerer_Leroy_2024}.
There are prospects for obtaining these much-needed measurements.
In the intermediate term, next-generation facilities that survey the full dust emission SED and the [\ion{C}{2}] line emission \citep[e.g., the proposed PRIMA mission;][]{Glenn_etal_2023} will lead to great advances.
In the near term, sub-mm surveys of dust emission \citep[with ground-based bolometers; e.g.,][]{Holland_etal_2013} and UV/optical observations of nebular and stellar attenuation are also viable paths forward \citep[e.g.,][]{Kreckel_etal_2013,Barrera_etal_2020,Faustino_etal_2024}, albeit with more methodological uncertainty.

For the CO emissivity term, a major issue is that dust-based, LVG-based, and carbon budget accounting methods yield incompatible $\alphaCO$ estimates in galaxy centers.
Given that these methods have been applied to many common targets, a comparative analysis that examines the input datasets and the assumptions underlying each method appears to be a fruitful next step.
A larger set of multi-line CO, $^{13}$CO, and [\ion{C}{1}] observations will further provide the necessary training data to calibrate CO emissivity prescriptions based on CO line width and other more physically relevant quantities accessible at high resolution \citep[e.g.,][]{Teng_etal_2024}.
Such observations are feasible, though expensive, with current facilities including ALMA.

For CO excitation, there is significant ongoing effort to measure $R_{21}$ and the related line ratios $R_{31}$, and $R_{32}$ across galaxies and spatial scales \citep[e.g.,][]{denBrok_etal_2021,Leroy_etal_2022,denBrok_etal_2023,Keenan_etal_2024a,denBrok_etal_2025,Komugi_etal_2025,Lee_etal_2025,Keenan_etal_2025}.
Recent studies highlighted $\SigSFR$ as the most promising predictor for $R_{21}$, and similar calibrations are beginning to emerge for the other line ratios \citep[][]{denBrok_etal_2024b,Keenan_etal_2025}.
Furthermore, simulations suggested that CO line ratios correlates with and may be used as proxies for CO emissivity variations \citep[e.g.,][]{Gong_etal_2020}.
In this sense, improved measurements and understanding of CO excitation effects may also feed back into improved prescriptions for the emissivity term.


\section{Conversion Factor Variations}
\label{sec:trends}

Based on our $\alphaCO$ predictions with the \citetalias{Schinnerer_Leroy_2024} prescriptions, we characterize $\alphaCO$ variations and trends across the local galaxy population (\autoref{sec:trends:resolved}--\ref{sec:trends:global}).
We are interested in how $\alphaCO$ depends on galactocentric radius, which captures much of its internal variation within galaxies.
We are also interested in the dependence of $\alphaCO$ on galaxy global stellar mass and SFR.
These are among the most fundamental properties in the context of galaxy evolution, and many galaxy-integrated CO surveys report key results as functions of $\Mstar$ and SFR \citep[e.g.,][]{Tacconi_etal_2013,Saintonge_etal_2017,Keenan_etal_2024a}.

\subsection{Conversion Factor Radial Profiles}
\label{sec:trends:resolved}

We first examine systematic variations in the resolved $\alphaCO$ radial profiles. 
Leveraging the large sample size, we compute median $\alphaCO$ radial profiles combining many galaxies with similar properties to average out galaxy-to-galaxy variations and distill key systematic trends.

\begin{figure}
\gridline{
\fig{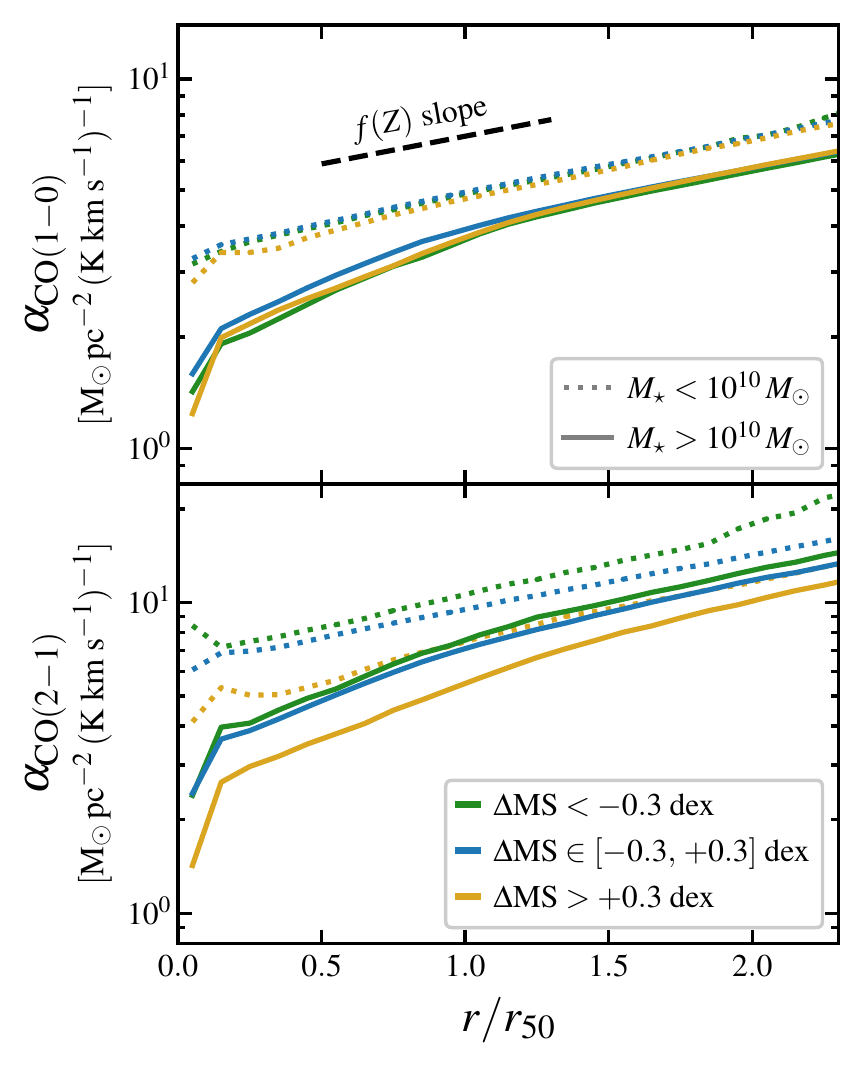}{0.48\textwidth}{}}
\vspace{-2.5\baselineskip}
\caption{Median $\alphaCO$ radial profiles for galaxies grouped by stellar mass ($\Mstar$) and offset from the star-forming main sequence ($\DeltaMS$; \autoref{eq:DeltaMS}). The $x$-axis represents galactocentric radius in units of $\radius{50}$ in WISE1 band. In lower-mass galaxies ($\Mstar<10^{10}\,\uM$, dotted lines), the radial profile of $\alphaCOxy10$ (\textit{top panel}) is effectively set by the CO-dark term $f(Z)$ alone. In higher-mass galaxies ($\Mstar>10^{10}\,\uM$, solid lines) the radial profile of $\alphaCOxy10$ steepens at $r \lesssim \radius{50}$ as the emissivity term $g(\Sigstar)$ plays a significant role. While the $\alphaCOxy10$ radial profiles barely vary with $\DeltaMS$ (color-coded), the $\alphaCOxy21$ radial profiles (\textit{bottom panel}) clearly do depend on this quantity. We predict lower $\alphaCOxy21$ in galaxies with higher $\DeltaMS$ (due to a higher line ratio $R_{21}$).}
\label{fig:radial}
\end{figure}

\autoref{fig:radial} shows median radial profiles of $\alphaCOxy10$ and $\alphaCOxy21$ for galaxies in various parts of the $\Mstar$--SFR parameter space.
These median radial profiles are derived for galaxies in bins of $\Mstar$ and ``offset'' from the star-forming main sequence (SFMS; also see \autoref{fig:SFMS} top left panel).
This offset is defined as
\begin{equation}
\DeltaMS = \log_{10}\SFR - \log_{10}\SFR\sbsc{MS}(\Mstar)~, \label{eq:DeltaMS}
\end{equation}
\noindent where $\mathrm{SFR}\sbsc{MS}(\Mstar)$ is the expected SFR at any given $\Mstar$ for galaxies on the SFMS.
Here we adopt a parameterized fit of the SFMS derived by \citet{Leroy_etal_2019} for the parent $z$0MGS sample:
\begin{equation}
\log_{10}\left(\frac{\SFR\sbsc{MS}}{\rm \uSFR}\right) = 0.68 \log_{10}\left(\frac{\Mstar}{10^{10}\,\uM}\right) - 0.17~.
\label{eq:SFMS}
\end{equation}
\noindent As such, galaxies with larger $|\DeltaMS|$ values are located further away from the SFMS (as defined by \autoref{eq:SFMS}) in the $\Mstar$--SFR plane.

\autoref{fig:radial} highlights several key trends.
First, galaxies in all $\Mstar$ and $\DeltaMS$ bins show strong radial gradients in $\alphaCOxy10$ and $\alphaCOxy21$, spanning at least a factor of 2 from $r=0$ to ${>}2\,\radius{50}$.
In relatively massive galaxies ($\Mstar > 10^{10}\,\uM$), the $\alphaCOxy10$ values can reach more than 3 times below the nominal Galactic value of $4.35\,\ualphaCO$ near galaxy centers, or more than 2 times above it in the outskirts (at $r \gtrsim 3\,\radius{50}$, outside the range of \autoref{fig:radial}).
This makes a clear case that accounting for $\alphaCO$ variations is important even when studying a single galaxy.

The median $\alphaCOxy10$ radial profiles for lower-mass galaxies ($\Mstar < 10^{10}\,\uM$) in all $\DeltaMS$ bins are consistent with a single exponential relation (i.e., a straight line under $\log(y)$ stretch), implying a 0.15~dex increase in $\alphaCOxy10$ when $r$ increases by $\radius{50}$.
This relation comes directly from the CO-dark term $f(Z)$, which combines a $-0.1\,\mathrm{dex}/\radius{50}$ metallicity gradient (see \autoref{sec:method:phys}) with a $Z^{-1.5}$ power-law scaling (\autoref{eq:CO-dark}).
The lack of any clear deviation from this simple expectation demonstrates that, for our fiducial prescription, the $\alphaCOxy10$ profiles are determined almost solely by the $f(Z)$ term at $\Mstar < 10^{10}\,\uM$.

The median $\alphaCOxy10$ radial profiles for higher-mass galaxies ($\Mstar > 10^{10}\,\uM$) have an overall lower normalization than less massive galaxies -- that is, $\alphaCOxy10$ appears lower at all given $r/\radius{50}$.
In addition, the $\alphaCOxy10$ profiles are no longer well described by a single exponential profile.
Instead, massive galaxies show clear deviations toward lower $\alphaCOxy10$ within $r < \radius{50}$.
This is due to common onset of the emissivity term $g(\Sigstar)$ near the centers of higher-mass galaxies, where $\Sigstar$ exceeds the $100\,\uSig$ threshold (see \autoref{eq:emissivity}).
Meanwhile, the overall $\alphaCOxy10$ profiles remain relatively insensitive to $\DeltaMS$ at $\Mstar > 10^{10}\,\uM$, reflecting weak variations in galaxy stellar mass distributions as a function of $\DeltaMS$.

The median $\alphaCOxy21$ radial profiles show a much stronger dependence on $\DeltaMS$, with systematically lower $\alphaCOxy21$ (at any given $r/r_{50}$) in galaxies with higher $\DeltaMS$.
In contrast to the lack of dependence seen for $\alphaCOxy10$, this clear $\DeltaMS$ dependence in $\alphaCOxy21$ comes from the line ratio term $R_{21}(\SigSFR)$.
As galaxies with higher $\DeltaMS$ generally have higher $\SigSFR$, $R_{21}$ should be higher and $\alphaCOxy21$ correspondingly lower.

\subsection{Galaxy Global Conversion Factors}
\label{sec:trends:global}

To connect more directly to the literature on \emph{global} galaxy evolution, we compute a global $\alphaCO$ value for each galaxy from the $\alphaCO$ radial profile via a WISE3 flux-weighted average.
That is, we use the WISE3 12~$\micron$ flux density in each radial bin (i.e., surface brightness times the radial bin area) as the weight to average $\alphaCO$ over all bins.
This averaging scheme is motivated by observational evidence of a strong and almost linear correlation between WISE3 surface brightness and CO line intensity \citep[e.g.,][]{Gao_etal_2019,Chown_etal_2021,Gao_etal_2022,Leroy_etal_2023a}.
The WISE3 flux-weighting thus approximates a CO flux-weighting, which is required to correctly convert radial profiles into global $\alphaCO$ values:
\begin{align}
\alphaCO\spsc{global}
&\equiv \frac{M\sbsc{mol,\,tot}}{L^\prime\sbsc{CO,\,tot}}
= \frac{\sum_i \alpha_{\mathrm{CO},i}\,I_{\mathrm{CO},i}\,A_i}{\sum_i I_{\mathrm{CO},i}\,A_i} \nonumber\\
&\approx \frac{\sum_i \alpha_{\mathrm{CO},i}\,I_{\mathrm{12\mu m},i}\,A_i}{\sum_i I_{\mathrm{12\mu m},i}\,A_i}~.
\label{eq:alphaCO_global}
\end{align}
\noindent Here $\alpha_{\mathrm{CO},i}$, $I_{\mathrm{CO},i}$, $I_{\mathrm{12\mu m},i}$, and $A_i$ denote the conversion factor (for either CO transition), CO line intensity (in $\uIco$), WISE3 surface brightness (in $\uI$), and area of the $i$th radial bin.
The last step is where the empirical finding of $I_\mathrm{12\mu m} \propto I_\mathrm{CO}$ comes in.

\begin{figure*}[htb]
\gridline{
\fig{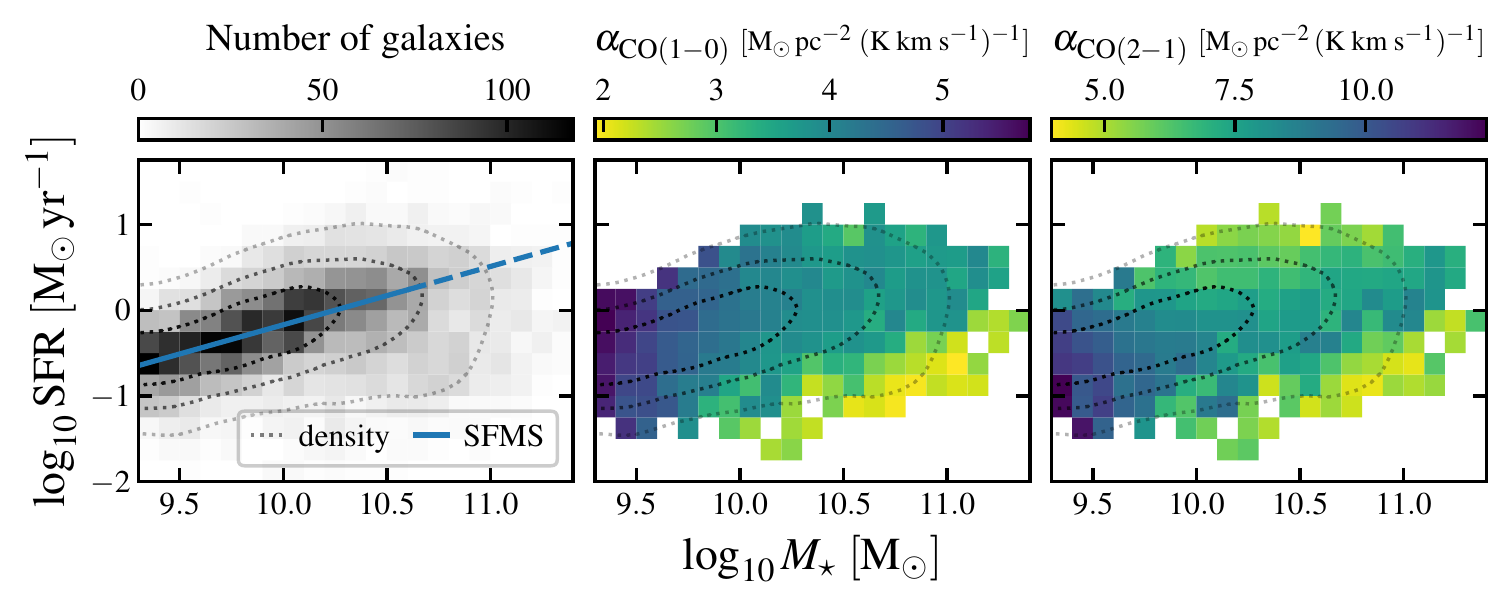}{\textwidth}{}}
\vspace{-2.5\baselineskip}
\gridline{
\fig{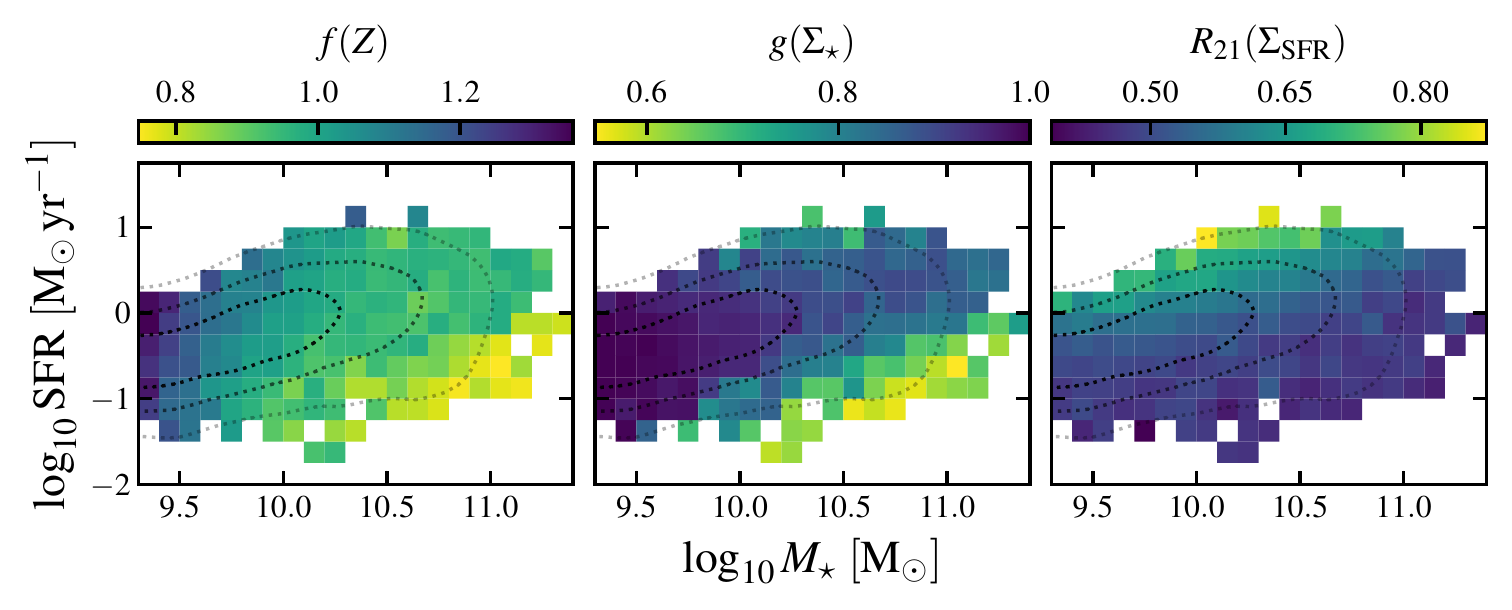}{\textwidth}{}}
\vspace{-2.5\baselineskip}
\caption{
Galaxy distribution and the predicted global $\alphaCO$ values across the $\Mstar$--SFR plane.
\textit{Top left:} 2D histogram (grayscale heatmap) and iso-density contours (dashed curves) of galaxies in our sample. Their distribution mostly centers around the star-forming main sequence at $z\approx0$ 
\citep[blue line;][]{Leroy_etal_2019}.
\textit{Top middle \& right:} Predicted galaxy global $\alphaCOxy10$ and $\alphaCOxy21$ values across the same plane (overlaid with the same density contours as in the top left panel).
The highest $\alphaCO$ values appear at the low $\Mstar$ end, and the lowest $\alphaCO$ values appear at higher $\Mstar$ and \emph{away} from the SFMS.
\textit{Bottom:} Variations of the CO-dark, emissivity, and line ratio terms in the \citetalias{Schinnerer_Leroy_2024} $\alphaCO$ prescriptions. These variations work in concert to drive the systematic trends seen in $\alphaCO$ in the top middle and right panels.
Note that the colorbar for $R_{21}$ (\textit{bottom right}) is reversed so that brighter colors still imply lower $\alphaCO$ as in other panels.
}
\vspace{0.3\baselineskip}
\label{fig:SFMS}
\end{figure*}

After obtaining global $\alphaCO$ values for all galaxies, we bin them into regular, logarithmically spaced grids of $\Mstar$ and SFR and report the results in the upper panels of \autoref{fig:SFMS}.
The upper left panel shows the distribution of all galaxies with $\alphaCO$ predictions.
As expected, most galaxies are located within $\pm$0.5~dex of the SFMS \citep[\autoref{eq:SFMS};][]{Leroy_etal_2019}, although the sample also includes a small subset of early-type galaxies reaching $\gtrsim$1~dex below the SFMS at the high $\Mstar$ end.

The upper middle and right panels of \autoref{fig:SFMS} show the variation of $\alphaCOxy10$ and $\alphaCOxy21$ across the $\Mstar$--SFR plane, with the color scale showing the median global $\alphaCO$ value in each $\Mstar$--SFR bin.
Both panels reveal systematic trends across the $\Mstar$ and SFR ranges probed by our sample.
Higher $\alphaCO$ values are generally seen towards lower $\Mstar$ for both CO transitions.
The lowest $\alphaCO$ values are found at intermediate to high $\Mstar$ \emph{away} from the SFMS -- for \CO10\ they mostly appear on the very low SFR extreme, whereas for \CO21\ they appear on both the low and high SFR extremes.

For most of these trends in the global $\alphaCO$ values visible in \autoref{fig:SFMS}, we can immediately identify corresponding trends in their resolved radial profiles in \autoref{fig:radial}.
For example, the lower global $\alphaCOxy10$ values in high-mass galaxies partly reflect the overall lower normalization of the $\alphaCOxy10$ radial profiles at high $\Mstar$ (\autoref{fig:radial} top panel); the lower global $\alphaCOxy21$ values in high-SFR galaxies instead originate from the $\DeltaMS$ dependence of the normalization of the $\alphaCOxy21$ radial profiles (\autoref{fig:radial} bottom panel).
In these cases, what we see in \autoref{fig:SFMS} are simply ``distilled'' versions of the systematic trends that we saw with the resolved $\alphaCO$ radial profiles.

But there is an intriguing trend in \autoref{fig:SFMS} that one would \emph{not} obviously expect from \autoref{fig:radial} --- the global $\alphaCOxy10$ values appear low in massive galaxies with low $\SFR$ or $\DeltaMS$.
This is in contrast to the \emph{insensitivity} of the resolved $\alphaCOxy10$ radial profiles to $\DeltaMS$ in \autoref{fig:radial}.
The reason behind this apparent contrast is the differential weighting of radial bins when deriving global $\alphaCO$ values (\autoref{eq:alphaCO_global}).
That is, massive galaxies with low $\DeltaMS$ tend to have more centrally concentrated WISE3 $12\,\micron$ emission, which means that the inner radial bins with lower $\alphaCO$ would receive more weight, resulting in lower global $\alphaCO$ values for these galaxies.
This interesting result is directly relevant to the interpretation of CO emission in lenticular and elliptical galaxies \citep[e.g.,][]{Young_etal_2011,Davis_etal_2019}.
Direct observational constraints on the CO emission radial distribution and conversion factors in these systems would help confirm the predicted trends and refine our understanding of the molecular gas properties therein.

As a further step in understanding and contextualizing the global $\alphaCO$ variations, we show in the lower panels of \autoref{fig:SFMS} systematic trends in the CO-dark term $f(Z)$, emissivity term $g(\Sigstar)$, and line ratio $R_{21}(\SigSFR)$ across the same $\Mstar{-}\SFR$ plane.
Here, the galaxy global averages of $f(Z)$, $g(\Sigstar)$, $R_{21}(\SigSFR)$ are computed with the same WISE3 flux-weighting scheme (similar to \autoref{eq:alphaCO_global} for $\alphaCO$).
The only exception is $R_{21}$, for which we compute a weighted \emph{harmonic mean} instead of arithmetic mean, reflecting the inverse proportionality between $\alphaCOxy21$ and $R_{21}$ (see \autoref{eq:alphaCO21}).

The lower left panel of \autoref{fig:SFMS} shows that the CO-dark term $f(Z)$ varies primarily with $\Mstar$, with the highest values appearing at the lowest $\Mstar$.
This trend is essentially ``baked in'' by our adopted galaxy mass--metallicity relation \citep{Sanchez_etal_2019} that sets the metallicity ``zero point'' for each galaxy (i.e., at $r=r_{50}$; see \autoref{sec:method:phys}).
Deviations from this dominant trend come from galaxy-to-galaxy variations in $r_{50}$ (which controls the radial metallicity slope; \autoref{sec:method:phys}) and the WISE3 surface brightness profile (which sets the relative weighting of all radial bins).
As mentioned above, the lower $f(Z)$ in high mass galaxies below the main sequence primarily reflects their more compact WISE3 distributions compared to galaxies with the same $M_\star$ on the star-forming main sequence.

The lower middle panel of \autoref{fig:SFMS} shows that the emissivity term $g(\Sigstar)$ is generally ``inactive'' at low $\Mstar$, as already discussed in \autoref{sec:trends:resolved}.
The behavior of $g(\Sigstar)$ at the high $\Mstar$ and low $\SFR$ end mirrors that of $\alphaCOxy10$ and $f(Z)$, and the explanation is essentially the same: more centrally concentrated WISE3 emission leads to high weights being assigned to inner radial bins with lower $g(\Sigstar)$ values.
These trends in $f(Z)$ and $g(\Sigstar)$ together drive the systematic variations of $\alphaCOxy10$ with $\Mstar$ and $\SFR$.

The lower right panel of \autoref{fig:SFMS} shows that the line ratio $R_{21}(\SigSFR)$ is mainly correlated with SFR or $\DeltaMS$.
This is consistent with the latest observations of galaxy global $R_{21}$ reported by \citet{Keenan_etal_2024} for galaxies with a similar range of $\Mstar$ and SFR.
More quantitatively, the median value in each SFR bin goes from $R_{21}=0.4$ at $\mathrm{SFR}=0.1\;\uSFR$ to $0.8$ at $10\;\uSFR$, also in good agreement with the results of \citet{Keenan_etal_2024}.
This monotonically increasing trend of $R_{21}$ with SFR and $\DeltaMS$ explains the difference between $\alphaCOxy10$ and $\alphaCOxy21$ seen in the upper middle and right panels of \autoref{fig:SFMS}.

\section{Implications for the Molecular Gas Depletion Time}
\label{sec:implications}

The molecular gas depletion time (defined as the ratio of molecular gas mass to star formation rate,  $\tdep\equiv\Mmol\,/\,\SFR$) and its dependencies on galaxy properties play a central role in our understanding of galaxy evolution \citep[e.g.,][]{Tacconi_etal_2020,Saintonge_Catinella_2022}.
In observational studies, $\tdep$ is often derived from the observable CO luminosity ($\LCO$) and SFR given some assumed $\alphaCO$ values:
\begin{equation}
\label{eq:sfrlco}
\tdep=\frac{\alphaCO\,\LCO}{\SFR}~.
\end{equation}
As such, obtaining reliable $\tdep$ estimates across galaxies requires understanding how $\alphaCO$ varies among the same galaxy population in the first place.

Our $\alphaCO$ estimates across a large sample of local galaxies allow us to address this problem statistically.
We use our $\alphaCO$ predictions to determine how $\SFR/\LCO$ \textit{should} depend on galaxy stellar mass $M_\star$ if \textit{only} $\alphaCO$ varies while $\tdep$ remains fixed.
Then we compare this prediction to the observed $\SFR/\LCO$ versus $M_\star$ trend for a large sample of literature observations.
The difference between the observed and predicted trends should reflect the real physical variations in $\tdep$ (or inaccuracies in our adopted $\alphaCO$ prescription).

\begin{figure*}[htb]
\gridline{
\fig{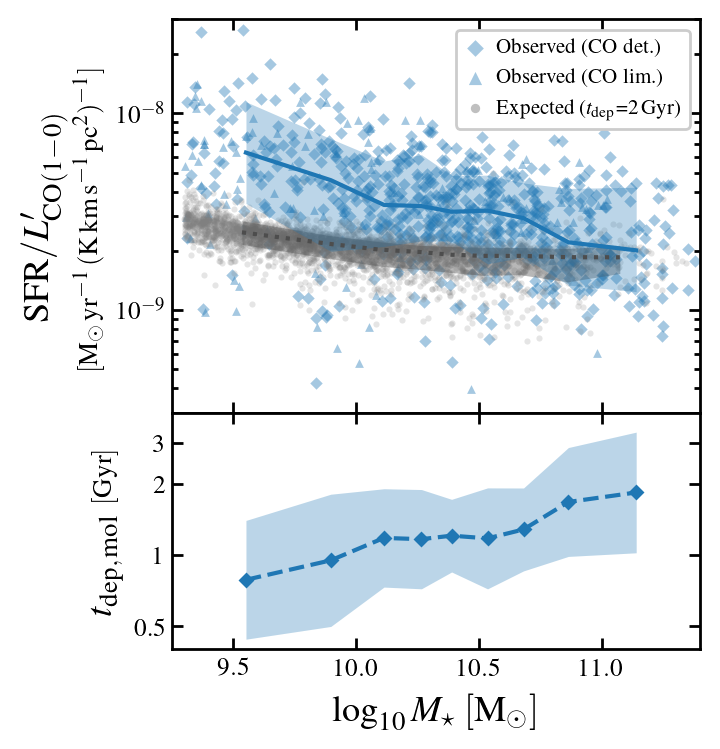}{0.49\textwidth}{}
\hfill
\fig{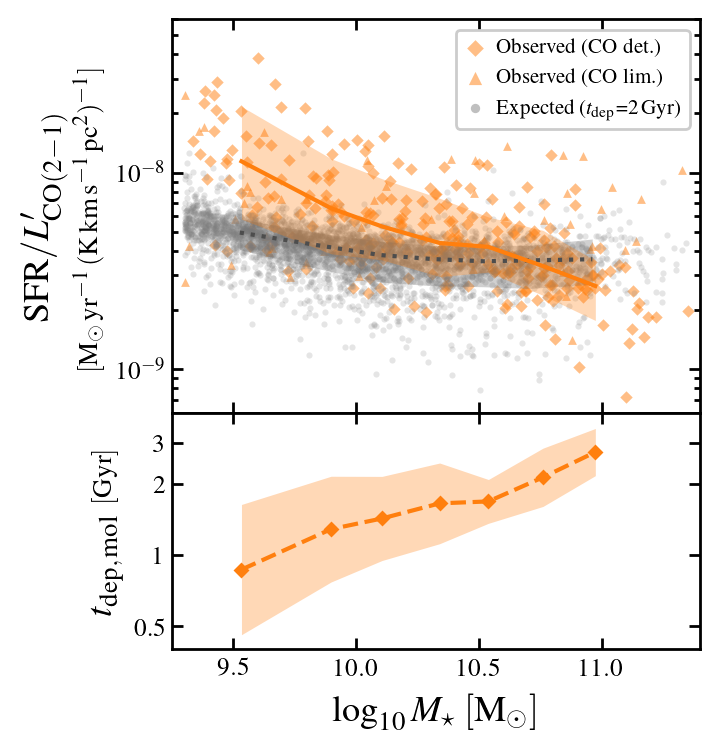}{0.49\textwidth}{}
}
\vspace{-2\baselineskip}
\caption{
Galaxy-integrated SFR to CO luminosity ratios as a function of stellar mass (\textit{top}) and the implied molecular gas depletion time trends (\textit{bottom}).
Results based on \CO10\ data are shown on the left, whereas those based on \CO21\ are on the right.
In the top panels, blue/orange symbols show a large set of observational measurements (diamonds) and CO upper limits (upward triangles, corresponding to $\SFR/\LCO$ lower limits) from the literature compiled by \citet{Leroy_etal_2022,Leroy_etal_2023a}.
The blue/orange solid lines show the running median for these observations, and shaded regions showing the 1$\sigma$ (16--84th percentile) range.
Gray symbols and dotted lines instead show the \emph{expected} relationship across our sample from our predicted global $\alphaCO$ values assuming a fixed molecular gas depletion time $t\sbsc{dep,\,fid}=2$~Gyr.
The discrepancies between the observations (blue/orange) and the expectations (gray) can be explained if $t\sbsc{dep}$ varies systematically as a function of $\Mstar$.
We show these implied variations in the bottom panels.
The implied $t\sbsc{dep}$ variations are $\sim$0.5~dex across the probed $\Mstar$ range and 0.1--0.3~dex at a given $\Mstar$.
}
\vspace{0.3\baselineskip}
\label{fig:SFRtoCO}
\end{figure*}

\vspace{0.5\baselineskip}
\textit{Observed $\SFR/\LCO$--$\Mstar$ relation:}
We consider $\SFR / \LCO$ vs. $\Mstar$ for an extensive compilation of integrated galaxy measurements from \citet{Leroy_etal_2022,Leroy_etal_2023a}. 
This compilation includes many surveys targeting either \CO10\ \citep[e.g., ALMaQUEST, AMIGA, EDGE-CARMA, xCOLDGASS;][]{Kuno_etal_2007,Lisenfeld_etal_2011,Bolatto_etal_2017,Saintonge_etal_2017,Sorai_etal_2019,LinLH_etal_2020,Wylezalek_etal_2022} or \CO21\ \citep[e.g., ALLSMOG, EDGE-APEX, HERACLES, PHANGS--ALMA;][]{Leroy_etal_2009,Bothwell_etal_2014,Jiang_etal_2015,Cicone_etal_2017,Colombo_etal_2020,Colombo_etal_2025,Leroy_etal_2021a}.
The $\Mstar$ and $\SFR$ values of all galaxies were derived with the same methods as $z$0MGS \citepalias{Leroy_etal_2019} and consistent with \citet{Salim_etal_2016,Salim_etal_2018}, so that the measurements for the literature and our sample are directly comparable to each other.
From this compilation, we select a subset of galaxies with $\Mstar{>}10^{9.3}\,\uM$ and within $\pm$0.5~dex of the SFMS (Figure \ref{fig:SFMS}) to focus on normal star-forming galaxies within the mass range where we predict $\alphaCO$.

We show the observed $\SFR/\LCO$ versus $\Mstar$ trends in the top panels of \autoref{fig:SFRtoCO}, plotting both CO detections (blue/orange diamonds) and upper limits (blue/orange upward triangles).
The running median curves for these observational data (blue/orange solid lines) show a systematic decrease in $\SFR/\LCO$ of 0.6--0.7~dex over ${\sim}1.5$~dex in $\Mstar$.
Both trends are reasonably described by a power law over the range of $M_\star = 10^{9.5{-}11}\,\uM$:
\begin{equation}
\frac{\SFR}{\LCOxy10} = \frac{10^{-8.37}\,\uSFR}{\uLco} \left(\frac{\Mstar}{10^{10}\,\uM}\right)^{-0.29}~,
\end{equation}
and
\begin{equation}
\frac{\SFR}{\LCOxy21} = \frac{10^{-8.19}\,\uSFR}{\uLco} \left(\frac{\Mstar}{10^{10}\,\uM}\right)^{-0.40}~.
\end{equation}
The scatters about these relations are $0.3{-}0.4$~dex.

\vspace{0.5\baselineskip}
\textit{Predicted $\SFR/\LCO$--$\Mstar$ relation:}
To construct a predicted trend, we select galaxies in our z0MGS sample within $\pm$0.5~dex of the SFMS (see \autoref{fig:SFMS}).
Then we combine their estimated $\alphaCO$ with an adopted fiducial depletion time $t\sbsc{dep,\,fid}=2\rm\;Gyr$ to predict $\SFR/\LCO$ for each galaxy via Equation \ref{eq:sfrlco}. 
This $t\sbsc{dep,\,fid}$ value is typical among resolved observations of massive, local, star-forming galaxies \citep[e.g.,][]{Leroy_etal_2008,Sun_etal_2023}, but the exact normalization is not critical because $t\sbsc{dep,\,fid}$ only scales the amplitude of $\SFR/\LCO$ for all galaxies, not its dependence on $\Mstar$.

We show the predicted $\SFR/\LCO$ as a function of $\Mstar$ for both CO transitions in the top panels of \autoref{fig:SFRtoCO}.
The galaxy-by-galaxy results (gray dots) and the running median curves (gray dotted lines) both reveal a weak decreasing trend of expected $\SFR/\LCO$ as a function of $\Mstar$.
In general, $\SFR/\LCO$ is predicted to decrease by 0.15--0.25~dex for either CO transition over ${\sim}1.5$~dex in $\Mstar$.
This reflects the anti-correlation between $\alphaCO$ and $\Mstar$ already seen in \autoref{fig:SFMS}.
The 1$\sigma$ scatter at any given $\Mstar$ is $\sim$0.1~dex.

\vspace{0.5\baselineskip}
\textit{Implications:}
The observational results shown in \autoref{fig:SFRtoCO} represent one of the most complete compilations to date for both \CO10\ and \CO21.
The observed trends are strong and continuous over the range of $M_\star = 10^{9.5{-}11}\,\uM$, where metallicity does not change by a large amount (for reference, the low end of this range corresponds to approximately the mass of M33 or the LMC).
Given that the observed trends in $\SFR/\LCO$ reflect a mixture of physical variations in $\alphaCO$ and $\tdep$, we suggest that its relationship with $\Mstar$ should be considered one of the fundamental molecular gas scaling relations.
We encourage future surveys to measure this relation explicitly and identify it as a valuable benchmark for numerical simulations that attempt to predict CO emission and star formation rate.

Furthermore, the \textit{predicted} trends from our $\alphaCO$ estimates and a fixed $\tdep$ is much weaker than the observed trends.
Observations show systematic changes of $0.6{-}0.7$~dex in $\SFR/\LCO$ as a function of $\Mstar$, while our calculations predict only $0.15{-}0.25$~dex over the same $\Mstar$ range.
Such discrepancies suggest that either our $\alphaCO$ predictions fail to match the amplitude of variations in reality, or $\tdep$ is not fixed in reality and varies systematically with $\Mstar$.

The most straightforward way to adjust our $\alphaCO$ prediction to explain the observations would be for $f(Z)$ to have a significantly steeper functional form, because metallicity varies primarily with $\Mstar$ in our calculations.
While not ruled out, this is not currently favored by simulation studies \citep[e.g.,][]{Gong_etal_2020,HuCY_etal_2022} or observational measurements in low-mass galaxies \citep[see][especially the synthesis plots in the latter]{Bolatto_etal_2013,Schinnerer_Leroy_2024}. 
For reference, $\alphaCO$ estimates for the LMC and M33, which would sit at the low end of our studied mass range, tend to find values of ${\sim}2$ times the Galactic value \citep[e.g.,][among others]{Leroy_etal_2011,Bolatto_etal_2013,Jameson_etal_2016}.
This translates to a ${\sim}0.3$~dex difference from value at the high-mass end, which is only half of the observed range in $\SFR / \LCO$.

Alternatively,  $\tdep$ may vary as a function of $\Mstar$, with lower values in low $\Mstar$ systems.
This appears to be the most likely explanation for our results.
By contrasting the observed trends with the predicted ones, we can derive the implied variations in $\tdep$ if our $\alphaCO$ predictions are accurate.
These implied $\tdep$ trends are shown in the bottom panels of \autoref{fig:SFRtoCO}.

In detail, the overall slope discrepancies suggest that $\tdep$ increases by $\sim$0.5~dex over $\Mstar=10^{9.5{-}11}\,\uM$, only reaching the fiducial 2~Gyr value near the high $\Mstar$ end.
The best-fit power-law relations are:
\begin{align}
\tdep\spsc{CO(1{-}0)} &= 1.2\,\mathrm{Gyr}\; (\Mstar/10^{10}\,\uM)^{0.19}~,\\
\tdep\spsc{CO(2{-}1)} &= 1.6\,\mathrm{Gyr}\; (\Mstar/10^{10}\,\uM)^{0.29}~.
\end{align}
The mild discrepancy between these trends inferred from \CO10\ and \CO21\ seems to suggest that our $R_{21}$ predictions may not fully account for its trends with $\Mstar$ in reality, or the \CO10\ and \CO21\ datasets in the literature may be influenced by different systematic effects, or both.

Another way to phrase our quantitative results is that roughly 1/3 of the observed trend in $\SFR/\LCO$ with $\Mstar$ are due to the predicted $\alphaCO$ variations, and the remaining 2/3 are due to $\tdep$ variations.
This is in reasonable agreement with results in the literature.
For example, \citet{Saintonge_etal_2017} reports an anti-correlation between $\tdep$ and $\Mstar$ based on the $\alphaCO$ prescription of \citet{Accurso_etal_2017}, and \citet{Hunt_etal_2020} find similar results based on their $\alphaCO$ estimates.

Exploring the reason for the shorter $\tdep$ in low-mass galaxies is beyond the scope of this paper.
However, we note that the ISM in these galaxies tends to be more dominated by \HI\ compared to \Htwo\ \citep[e.g.,][]{Saintonge_Catinella_2022}, they tend to host fewer molecular clouds with lower surface densities \citep[e.g.,][]{Sun_etal_2022}.
The shorter $\tdep$ in low-$\Mstar$ galaxies may reflect that these galaxies lack diffuse or inert molecular gas, that atomic gas makes up more of the material in star-forming complexes, and/or that feedback more readily disperses molecular clouds.


\section{Summary}
\label{sec:summary}

We analyze GALEX UV and WISE IR images for \ngalall\ local galaxies to measure the UV and IR band surface brightness radial profiles, half-light radii, stellar mass surface density ($\Sigstar$) profiles, and SFR surface density ($\SigSFR$) profiles for those detected in the relevant bands.
This represents a major effort to systematically characterize the radial structure of massive galaxies ($M_\star>10^{9.3}\,\uM$) throughout the local volume ($d\lesssim50$~Mpc), as part of the $z{=}0$ Multiwavelength Galaxy Synthesis project \citep{Leroy_etal_2019}.

While these measurements have broad relevance and many potential applications, in this work we primarily use them to obtain CO-to-H$_2$ conversion factor ($\alphaCO$) predictions based on state-of-the-art empirical prescriptions \citep[e.g.,][]{Bolatto_etal_2013,Gong_etal_2020,Schinnerer_Leroy_2024}.
These spatially resolved $\alphaCO$ predictions cover \ngalxco\ galaxies, many of which have been (or will likely be) targeted in large CO mapping campaigns.
Our $\alphaCO$ predictions address the urgent need for modern CO studies to account for substantial $\alphaCO$ variations both within a galaxy and among galaxies.

To validate our $\alphaCO$ predictions, we compile existing observational $\alphaCO$ measurements in the literature \citep[including dust-based, CO multi-line modeling, and carbon budget accounting results;][]{Sandstrom_etal_2013,Israel_2020,denBrok_etal_2023,Teng_etal_2023,Yasuda_etal_2023,Chiang_etal_2024}, and check our predictions against these measurements for a small overlapping subsample of galaxies (\autoref{fig:alphaCO10} and \autoref{fig:alphaCO21}).
The best agreements are achieved when the predictions account for not only CO-dark gas (as a function of metallicity), but also CO emissivity (as a function of $\Sigstar$) and excitation effects (as a function of $\SigSFR$).
The emissivity effects are crucial for properly reproducing the observed low $\alphaCO$ values in the inner regions of massive galaxies, although there remains some tension between different observational measurements in this regime.
The excitation effects play a central role in translating between CO transitions and should be considered when using \CO21\ or higher-$J$ lines to trace molecular gas.

Across our entire sample, the $\alphaCO$ predictions exhibit several salient trends as functions of galaxy global stellar mass ($\Mstar$) and SFR (\autoref{fig:radial} and \autoref{fig:SFMS}).
For low-$\Mstar$ galaxies, the metallicity-dependent CO-dark term dominates the hybrid \citet{Schinnerer_Leroy_2024} prescription, predicting high values of $\alphaCOxy10\gtrsim5$ and $\alphaCOxy21\gtrsim10$ (in units of $\ualphaCO$).
For high-$\Mstar$ but low-SFR galaxies, the $\Sigstar$-dependent emissivity term becomes prominent and predicts low values of $\alphaCOxy10\lesssim2.5$ and $\alphaCOxy21\lesssim5$.
For high-SFR galaxies, the $\SigSFR$-dependent excitation term predicts high \CO21-to-(1--0) ratios of $R_{21}\gtrsim0.8$, which suggests lower $\alphaCOxy21$ values even though $\alphaCOxy10$ may be moderate.

We explore the implications of our prediction for the molecular gas depletion time, $\tdep$ (\autoref{fig:SFRtoCO}).
Leveraging an extensive compilation of galaxy global SFR and CO luminosity measurements, we measure the dependence of $\SFR/\LCO$ on $\Mstar$ for main sequence galaxies to be $\SFR/\LCOxy10 \propto \Mstar^{-0.29}$ and $\SFR/\LCOxy21 \propto \Mstar^{-0.40}$.
We contrast these measurements with expectations from our $\alphaCO$ predictions and a fixed $\tdep$, finding that the observed trends are much stronger than the expected trends and exhibit ${\sim}3$ times wider range of $\SFR/\LCO$ values over the same $\Mstar$ range.
The most likely explanation is that the molecular gas depletion time increases systematically with $\Mstar$, from ${\lesssim}1$~Gyr at $\Mstar=10^{9.5}\,\uM$ to 2--3~Gyr at $10^{11}\,\uM$.

To facilitate many potential applications of our results, we publish all data products from this work on the Canadian Advanced Network for Astronomical Research (CANFAR)\footnote{A live copy of these data products is available at \href{https://www.canfar.net/storage/vault/list/z0MGS/Sun_etal_2025/}{this URL}.}.
These products include resolved radial profiles and galaxy integrated properties (in the form of machine-readable tables, see \autoref{sec:trends:global}) as well as $\alphaCO$ predictions for subsamples of galaxies in several CO surveys (in the form of FITS images, see \autoref{apdx:maps}).
We also publish a \texttt{Python} package that compiles and implements many empirical $\alphaCO$ prescriptions in the literature (\autoref{apdx:python}).
We hope that these efforts will motivate the community to adopt current best practices for handling $\alphaCO$, especially in the context of large CO surveys \citep[e.g., EDGE-CALIFA, MAUVE--ALMA, and KILOGAS;][J.~Sun et al., in preparation]{Bolatto_etal_2017}.
We also anticipate future works to expand and improve the current set of $\alphaCO$ measurements from observations.


\vspace{\baselineskip}
{

JS acknowledges support by the National Aeronautics and Space Administration (NASA) through the NASA Hubble Fellowship grant HST-HF2-51544 awarded by the Space Telescope Science Institute (STScI), which is operated by the Association of Universities for Research in Astronomy, Inc., under contract NAS~5-26555.
YHT and ADB acknowledge support from grant NSF-AST 2307441.
AKL and RC gratefully acknowledge support from NSF-AST 2205628, JWST-GO-02107.009-A, and JWST-GO-03707.001-A. AKL also gratefully acknowledges support by a Humboldt Research Award.
AH acknowledges support by the Programme National Cosmology et Galaxies (PNCG) of CNRS/INSU with INP and IN2P3, co-funded by CEA and CNES, and by the Programme National Physique et Chimie du Milieu Interstellaire (PCMI) of CNRS/INSU with INC/INP co-funded by CEA and CNES.

This work uses data from the \textit{Galaxy Evolution Explorer (GALEX)}. \textit{GALEX} is a NASA Small Explorer, whose mission was developed in cooperation with the Centre National d'Etudes Spatiales (CNES) of France and the Korean Ministry of Science and Technology. \textit{GALEX} is operated for NASA by the California Institute of Technology under NASA contract NAS5-98034.

This work uses data from the \textit{Wide-field Infrared Survey Explorer (WISE)}, which is a joint project of the University of California, Los Angeles, and the Jet Propulsion Laboratory/California Institute of Technology, funded by NASA.

The z0MGS project and the creation of the GALEX and WISE atlas was supported by NASA ADAP grants NNX16AF48G and NNX17AF39G and National Science Foundation grant No.~1615728.
The \textit{GALEX} and \textit{WISE} data used in this paper can be found in the NASA/IPAC Infrared Science Archive: \dataset[10.26131/IRSA6]{http://dx.doi.org/10.26131/IRSA6}.

We acknowledge the usage of the SAO/NASA Astrophysics Data System and the HyperLEDA database.

\facilities{
GALEX, WISE, ALMA, APEX, ARO:12m, ARO:SMT, CARMA, IRAM:30m, FCRAO, No:45m
}

\software{
\texttt{NumPy} \citep{NumPy_2020},
\texttt{SciPy} \citep{SciPy_2020},
\texttt{Matplotlib} \citep{Matplotlib_2007},
\texttt{Astropy} \citep{Astropy_2013,Astropy_2018,Astropy_2022},
\texttt{APLpy} \citep{APLpy_2012},
\texttt{CO\_conversion\_factor} (\url{https://github.com/astrojysun/COConversionFactor}),
\texttt{adstex} (\url{https://github.com/yymao/adstex}).
}

}


\appendix
\restartappendixnumbering
\twocolumngrid

\section{Stripe Integral}
\label{apdx:stripe}

We use a ``stripe integral'' method \citep{Warmels_1988} to reconstruct surface brightness radial profiles for galaxies with high inclination angles (see \autoref{sec:method:radprof}).
In this appendix, we detail the concept behind this method, our implementation, and some caveats.

\begin{figure}[tb]
\gridline{
\fig{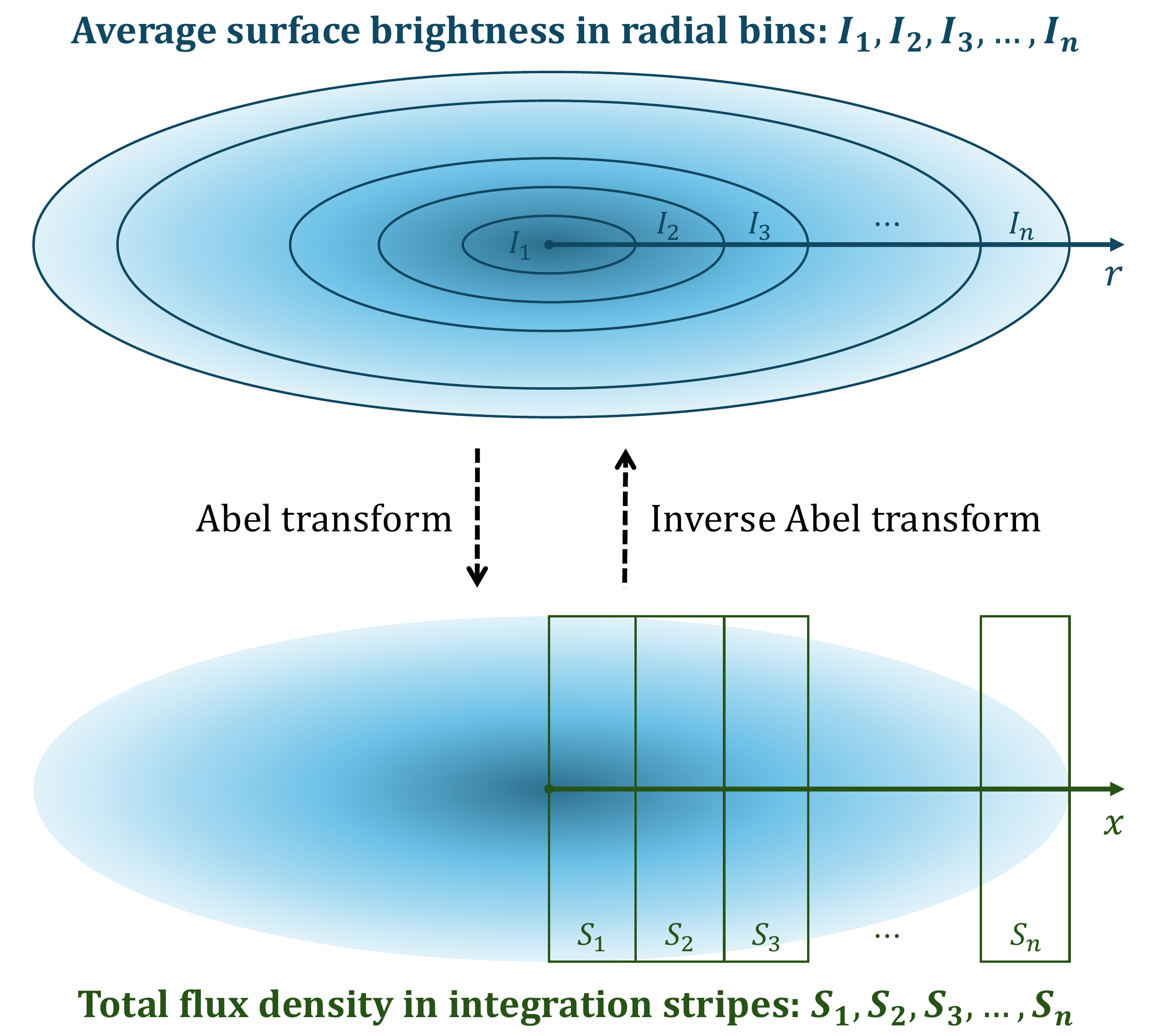}{0.48\textwidth}{}
}
\vspace{-2\baselineskip}
\caption{
Illustration of a surface brightness profile measured from radial binning ($I_1, I_2, ..., I_n$; \textit{top}) and a flux density sequence computed via stripe integrals ($S_1, S_2, ..., S_n$; \textit{bottom}).
For axisymmetric disks, they are related by a discrete version of the Abel transform (or its inverse).
}
\vspace{0.3\baselineskip}
\label{fig:stripe_int}
\end{figure}

\autoref{fig:stripe_int} illustrates the concept with an inclined galaxy disk that is geometrically thin, axisymmetric, and has optically thin emission.
This distribution, when viewed face-on, follows a surface brightness radial profile, $I(r)$.
A common way to derive $I(r)$ for moderately inclined galaxies is to calculate the average observed surface brightness in a series of radial bins and multiply them by $\cos{i}$ to account for the inclination.
This is the ``radial binning'' method, and the result would be a series of inclination-corrected surface brightnesses ($I_1, I_2, ..., I_n$) as a discrete representation of the true surface brightness profile $I(r)$.

The radial binning method works under three conditions: (a) the inclination and position angle of the galaxy are well known, (b) the light distribution along the (projected) minor axis is well resolved in observations, and (c) the intrinsic ``flatness'' of the galaxy disk, defined as the disk vertical scale height to radial scale length ratio, is much smaller than $\cos{i}$.
For high-inclination galaxies that do not fulfill these conditions, there will be substantial ``smearing'' of emission along the minor axis across radial bins, such that the radial binning measurements become at least inaccurate, if not completely meaningless.
In the case where the inclination is poorly constrained, the conversion to face-on surface brightness estimates will also be highly uncertain.

The stripe integral method reconstructs $I(r)$ even in cases where these conditions do not hold. 
Instead of radial bins, it defines a series of rectangular ``integration stripes,'' each spanning the full width of the galaxy along its minor axis (see bottom part of \autoref{fig:stripe_int}).
The width of these integration stripes along the major axis matches the desired radial bin width for the output surface brightness profile, and the stripes collectively cover the entire galaxy footprint.
Within each stripe, one computes the enclosed total flux density, $S=\int I d\Omega$.
This sequence of integrated flux densities across all stripes ($S_1, S_2, ..., S_n$) represents a key intermediate product.

Under the assumptions of an axisymmetric disk, optically thin emission, and no emission beyond the last radial bin, one can translate between the radial surface brightness profile $I_1, I_2, ..., I_n$ and the flux density sequence $S_1, S_2, ..., S_n$ via a discrete version of the Abel transform (or its inverse transform).
For such linear transforms between $n$-dimensional vectors, one can write them in matrix form as:
\begin{equation}
\begin{bmatrix}
S_1 \\ S_2 \\ \vdots \\ S_n
\end{bmatrix}
= A
\begin{bmatrix}
I_1 \\ I_2 \\ \vdots \\ I_n
\end{bmatrix}~,\qquad\text{or}~
\begin{bmatrix}
I_1 \\ I_2 \\ \vdots \\ I_n
\end{bmatrix}
= A^{-1}
\begin{bmatrix}
S_1 \\ S_2 \\ \vdots \\ S_n
\end{bmatrix}~.
\end{equation}
\noindent Here, $A$ is the matrix representation of the Abel transform.
Its matrix elements $A_{ij}$ can be interpreted geometrically as the (deprojected) overlapping area between the $i$th integration stripe and the $j$th radial bin.
Given the number of stripes/bins and their width in angular units, one can calculate all the matrix elements according to this geometric interpretation.
One can then compute the inverse of $A$ and use it to convert the flux density sequence $S_1, S_2, ..., S_n$ to the radial surface brightness profile $I_1, I_2, ..., I_n$.

We note that this matrix-based implementation differs from the iterative approach adopted by \citet{Warmels_1988}.
There, the radial surface brightness profile is solved iteratively using the Richardson--Lucy algorithm \citep{Richardson_1972,Lucy_1974}.
While this algorithm has desirable features such as enforcing nonnegative surface brightness and remaining robust against noise, its computational cost becomes non-negligible when applied to thousands of galaxies with six-band images at different resolutions.
Our matrix-based implementation is much less computation-intensive and thus more practical when dealing with a large galaxy sample, as is the case with this study.

To account for the effect of noise, we calculate the measurement errors on the stripe-integrated flux densities $S_1, S_2, ..., S_n$, propagate them to the output surface brightness profile $I_1, I_2, ..., I_n$, and truncate the profile where the S/N ratio per bin drops below 3.
The effect of noise accumulates from large to small galactocentric radius in this setup.
As a result, the surface brightness profile derived from stripe integral often drops below $\text{S/N}=3$ at smaller radii compared to that derived from direct radial binning for low-inclination galaxies.

Another notable caveat when applying the stripe integral method to our data is the presence of pixels masked due to the presence of foreground stars or background galaxies (see \autoref{sec:method:data}) and outlier pixels (e.g., unmasked stars).
For masked pixels, we cannot simply assign them the mean or median of the unmasked pixels in a stripe because we do not expect those pixels to have similar surface brightnesses (this differs from the case of radial binning).
For unmasked stars, the effect of these outliers can accumulate and end up affecting the output surface brightness profile across all inner radial bins.

To address these issues, we prepare the images in the following way before calculating the stripe integral: for each pixel in the image, we find its reflection about the galaxy center, major axis, and minor axis, and use the median among these four pixels (ignoring masked pixels) to replace the original pixel value.
This step leverages the four-fold symmetry about the galaxy axes expected for axisymmetric galaxy disks at all inclination angles.
It allows for filling in masked pixels in a reasonable way and reduces the impact of outlier pixels on the stripe integral calculation because they contribute only 1 of 4 points in the median. 
For moderately inclined galaxies, we generally see better agreement between the stripe integral-based results and radial binning after implementing this preparatory step, which confirms its effectiveness.

\section{Resolution Effects}
\label{apdx:res}

\begin{figure*}[htb]
\gridline{
\fig{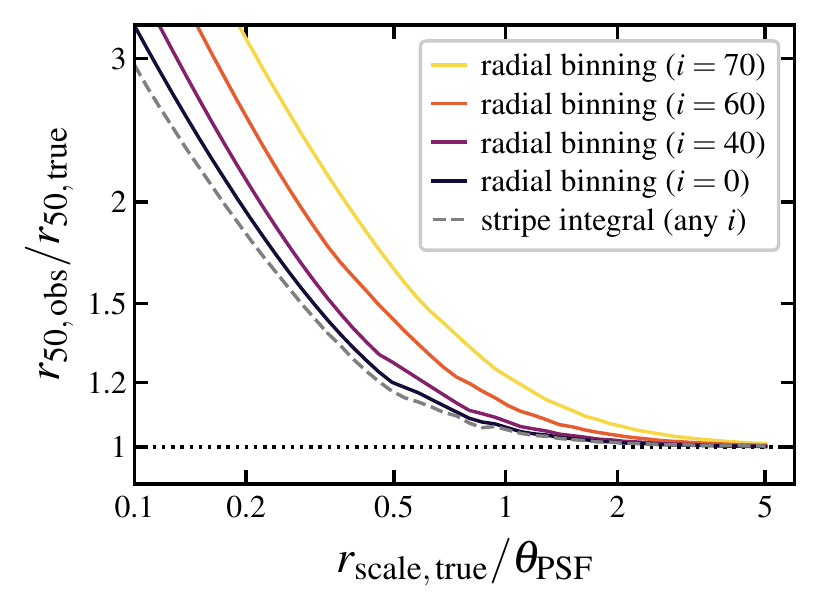}{0.48\textwidth}{}
\hfill
\fig{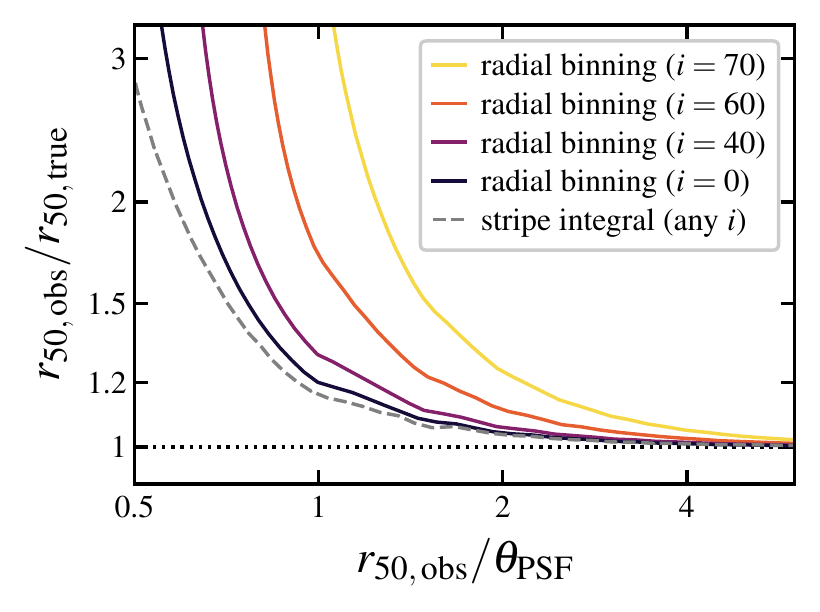}{0.48\textwidth}{}
}
\vspace{-2.5\baselineskip}
\caption{
Effects of finite data resolution on galaxy half-light radius ($\radius{50}$) measurements.
The $y$-axes show the discrepancy between the observed half-light radius ($\radius{50,\,obs}$) and the true value ($\radius{50,\,true}$), which diminishes as the galaxy becomes better resolved (towards larger $x$ values).
The $x$-axis represents the ratio of galaxy exponential scale radius ($\radius{scale,\,true}$) over PSF size ($\theta\sbsc{PSF}$) in the left panel, or the ratio of the observed effective radius ($\radius{50,\,obs}$) over PSF size in the right panel.
When measuring $\radius{50}$ with the direct radial binning method (see \autoref{sec:method:radprof}), resolution-induced discrepancies are more substantial for more inclined galaxies (brighter colored lines); in contrast, $\radius{50}$ measured via the stripe integral method (see \autoref{sec:method:radprof} and \autoref{apdx:stripe}) is insensitive to galaxy inclination.
Results shown by the right panel can be used to correct for resolution-induced biases in $\radius{50}$, as both $\radius{50,\,obs}/\theta\sbsc{PSF}$ ($x$-axis) and $i$ (color-code) are directly measurable from observations.
}
\vspace{0.3\baselineskip}
\label{fig:res}
\end{figure*}

One of the key intermediate measurements in this work is the galaxy half-light radius, $\radius{50}$, which we derive from the radial surface brightness profile for each band (see \autoref{sec:method:radprof:r50}).
This measurement can be affected by the finite resolution of the images, especially when the target galaxy is small and its radial profile is not fully resolved.
In this appendix, we assess this effect by analyzing images of mock galaxies with varying sizes and inclination angles and quantifying the systematic biases due to resolution limit.

We create a set of mock images of galaxies.
Each model is an axisymmetric disk with an exponential surface brightness profile, but with varying galaxy sizes and inclination angles.
The ratio of the galaxy size, determined by the exponential scale length ($\radius{scale,\,true}$), to the PSF size ($\theta\sbsc{PSF}$), determines how well the galaxy is resolved.
We define a grid that spans $\radius{scale,\,true}/\theta\sbsc{PSF}=0.1{-}5.0$, from unresolved to well resolved.
Our grid of inclinations spans $i=0^\circ{-}85^\circ$, i.e., from face-on to almost edge-on.

For each combination of $\radius{scale,\,true}/\theta\sbsc{PSF}$ and $i$, we project a disk with an exponential surface brightness profile onto the sky plane and convolve it with the appropriate Gaussian PSF to create a mock image.
We then measure its surface brightness profile via both the radial binning and stripe integral approaches, and so derive the half-light radius ($\radius{50,\,obs}$) in the same way as we would for a real galaxy (see \autoref{sec:method:radprof}).
By comparing the observed $\radius{50,\,obs}$ and true value $\radius{50,\,true}\approx1.68\,\radius{scale,\,true}$ for each mock galaxy, we measure the systematic bias imposed by finite resolution.

The left panel in \autoref{fig:res} shows the results of these experiments.
As expected, the ratio of observed to true galaxy size ($\radius{50,\,obs}/\radius{50,\,true}$) is close to unity in the well-resolved regime, i.e., $\radius{scale,\,true}/\theta\sbsc{PSF}>2$.
The galaxy size becomes increasingly overestimated at smaller $\radius{50,\,true}/\theta\sbsc{PSF}$, reaching a factor of 2 at $\radius{scale,\,true}/\theta\sbsc{PSF}\approx0.2{-}0.4$.
The colored lines show a secondary dependence on the inclination angle when measuring the surface brightness profile via radial binning.
As expected, we observe no dependence of $\radius{50,\,obs}/\radius{50,\,true}$ on the inclination when inferring $\radius{50,\,obs}$ using the stripe integral technique.

Based on these results, we correct our galaxy size measurements for resolution-induced bias. 
To do this, we first recast our model $(\radius{scale,\,true}/\theta\sbsc{PSF},\,i)$ grid into a $(\radius{50,\,obs}/\theta\sbsc{PSF},\,i)$ grid, in which both parameters can be measured from observations.
This is done by calculating $\radius{50,\,obs}/\theta\sbsc{PSF}=(\radius{50,\,obs}/\radius{50,\,true}) \times (1.68\,\radius{scale,\,true}/\theta\sbsc{PSF})$ for each node in the model grid.
Then we interpolate the $\radius{50,\,obs}/\radius{50,\,true}$ values from the model grid nodes (see \autoref{fig:res} right panel) to cover the corresponding $(\radius{50,\,obs}/\theta\sbsc{PSF},\,i)$ space spanned by our grid.
At high $\radius{50,\,obs}/\theta\sbsc{PSF} > 5$ i.e., outside our measured grid, we treat $\radius{50,\,obs}$ as the true value, since virtually no correction is required. 
At low $\radius{50,\,obs}/\theta\sbsc{PSF}$, we truncate the interpolated values at $\radius{50,\,obs}/\radius{50,\,true}=2$, beyond which this correction factor becomes too large and too sensitive to small variations in the input parameter.
In those cases, we deem it impossible to robustly recover the true galaxy size $\radius{50,\,true}$ from the observations.

\section{Conversion Factor Maps}
\label{apdx:maps}

We provide two-dimensional predicted $\alphaCO$ maps for several modern CO surveys, including 
COMING \citep{Sorai_etal_2019}, 
HERACLES \citep{Leroy_etal_2009}, 
PHANGS--ALMA \citep{Leroy_etal_2021a},
as well as VERTICO \citep{Brown_etal_2021} and its high-resolution successor MAUVE--ALMA (J.~Sun et al., in preparation).
Details specific to each sample can be found in the bullet points below.
To maximize the utility of these maps, we refine the data processing for each sample by (1) updating key galaxy parameters (e.g., inclination and position angles) to match the adopted values in the corresponding survey papers, (2) matching the WCS grids of the $\alphaCO$ maps to the CO maps published by each survey, and (3) visually inspecting all products and making additional adjustments when necessary.

\begin{itemize}[itemsep=0.2em,leftmargin=1em,parsep=0.3em,partopsep=-0.1em]

\item[$\sbullet$] COMING:
We use galaxy central coordinates, inclination, and position angles adopted in \citet[Table~1]{Sorai_etal_2019}, and stellar mass from the $z$0MGS catalog.
Among all COMING targets, we omit all interacting galaxies and one other galaxy (IC~10) that is not in the $z$0MGS sample.
For the remaining 120 galaxies, we derived their radial profiles and half-light radii in a similar way as described in Section~\ref{sec:method}.
Eleven galaxies were flagged in this step due to high fractions of masked area in their WISE1 images (Section~\ref{sec:method:data}).
Finally, we project the estimated $\alphaCOxy10$ radial profile onto the WCS grids of the COMING \CO10\ maps.
This last step is done only for low-inclination galaxies ($i < 75^\circ$), because rigorous treatments of high-inclination galaxies require reconstructing the actual radial profile of CO emission rather than naively projecting the $\alphaCO$ radial profiles onto the sky plane.

\item[$\sbullet$] HERACLES:
We consider all 48 HERACLES targets, including the ones appeared in \citet{Leroy_etal_2013}.
We use the central coordinates, inclination and position angles, as well as stellar mass and size measurements later compiled as part of the analyses presented in \citet{Leroy_etal_2023a}.
For estimating the $\alphaCO$ radial profiles, we omit 16 galaxies with stellar mass below $10^{9.3}\,\uM$ (see Section~\ref{sec:method}) and two more (NGC~6946 and NGC~7331) due to high fractions of masked area.
All remaining targets have low inclinations so that we can directly project their $\alphaCOxy21$ profiles onto the WCS grids of the published HERACLES \CO21\ maps.

\item[$\sbullet$] PHANGS--ALMA:
We use the central coordinates, inclination and position angles, stellar masses, and sizes adopted in \citet[Tables~3 \& 4]{Leroy_etal_2021a}.
For the radial profile analyses, we omit twelve galaxies with high fractions of masked area in WISE1.
We produce $\alphaCOxy21$ maps matching the WCS grids of the published PHANGS--ALMA \CO21\ data for all remaining galaxies except NGC~4207 (which likely has a higher inclination angle than the adopted value).

\item[$\sbullet$] VERTICO \& MAUVE--ALMA\footnote{The MAUVE--ALMA survey (J.~Sun et al., in preparation) targets 40 out of 51 VERTICO galaxies in \CO21. Therefore, the $\alphaCOxy21$ maps created here can be used for both surveys.}:
We use the central coordinates, inclination, and position angles adopted in \citet[Table~1]{Brown_etal_2021}, as well as stellar mass from the $z$0MGS catalog.
Among the 51 VERTICO targets, we omit two galaxies without any CO detection (IC~3418 and VCC~1581), two interacting galaxies (NGC~4567/8), and flag one more galaxy (NGC~4561) with a high fraction of masked area in WISE1.
After deriving all radial profiles, we make $\alphaCOxy21$ maps matching the WCS grids of the VERTICO \CO21\ data for all low-inclination galaxies.

\end{itemize}

We also considered other local galaxy CO surveys such as AlFoCS \citep{Zabel_etal_2019}, ALMaQUEST \citep{LinLH_etal_2020}, CARMA--EDGE \citep{Bolatto_etal_2017}, and the Fornax ACA survey \citep{Morokuma-Matsui_etal_2022}.
However, the overlaps between these samples and ours are limited due to mismatches in the sample selection criteria, e.g., distance range (for ALMaQUEST and CARMA--EDGE) or stellar mass range (for the Fornax surveys).
Since the sample selection of $z0$MGS is driven by the resolution and sensitivity of WISE, this implies that other data sets are likely better suited to underpin conversion factor estimates for those surveys.
Therefore, we consider that creating conversion factor maps for these samples is beyond the scope of this work and more suitable for dedicated follow-up efforts.

\section{Python Implementation of Conversion Factor Prescriptions}
\label{apdx:python}

We provide \texttt{CO\_conversion\_factor}, a \texttt{Python} package that is \texttt{pip}-installable\footnote{\url{https://pypi.org/project/CO-conversion-factor/}} and compiles many existing CO-to-\Htwo\ conversion factor prescriptions in the literature.
In addition to the prescriptions used in this work \citep{Bolatto_etal_2013,Gong_etal_2020,Schinnerer_Leroy_2024}, the compilation also includes \citet{Narayanan_etal_2012}, \citet{Amorin_etal_2016}, \citet{Accurso_etal_2017}, \citet{Sun_etal_2020b}, and \citet{Teng_etal_2024}.
Each prescription (or family of prescriptions) is implemented as a stand-alone \texttt{Python} function that takes a set of input parameters relevant to the prescription (such as metallicity or gas mass surface density) and returns its conversion factor prediction.
Note that these prescriptions are often calibrated within a finite range of input parameters.
Their applications should therefore be limited to within the same range of parameter space.

Since metallicity is the most common input parameter for these prescriptions, but the coverage of high-quality metallicity measurements among nearby galaxies remains scarce, the package also includes the scaling relation-based metallicity prescriptions used in \citet{Sun_etal_2020b,Sun_etal_2020a,Sun_etal_2022,Sun_etal_2023} and this work (\autoref{sec:method:phys}).
These prescriptions leverage galaxy mass-metallicity relations and metallicity radial gradients measured from large surveys \citep[e.g.,][]{Sanchez_etal_2014,Sanchez_etal_2017,Sanchez_etal_2019}.
We recommend using these prescriptions only when direct observational measurements are unavailable.



\bibliography{main}
\bibliographystyle{aasjournal}




\end{document}